\documentclass[prx,twocolumn,english,superscriptaddress,floatfix,longbibliography]{revtex4-2}

\usepackage{natbib}
\usepackage{physics}
\usepackage{xcolor}
\usepackage{amsmath, amsfonts}
\usepackage{siunitx}
\usepackage{graphicx}
\usepackage[english]{babel}
\usepackage[utf8]{inputenc}
\usepackage[autostyle]{csquotes}
\MakeOuterQuote{"}
\usepackage[shortlabels]{enumitem}
\usepackage{dsfont}
\usepackage{comment}
\usepackage{verbatim}
\usepackage{xpatch}
\usepackage{tcolorbox}
\usepackage{float}
\usepackage{etoolbox}
\usepackage[normalem]{ulem}

\usepackage[hidelinks]{hyperref}	
\usepackage[capitalize]{cleveref}
\date{\today}

\newcommand{\set}[1]{\{#1\}}

\newcommand{\PP}[1]{{\hat{P}\left[#1\right]}}

\definecolor{darkgreen}{rgb}{0.0, 0.8, 0.0}

\usepackage{etoolbox}

\definecolor{pink}{RGB}{255,0,255}

\newcommand{\affvqcc}{Vigo Quantum Communication Center, University of Vigo, Vigo E-36310, Spain}
\newcommand{\affuvigo}{Escuela de Ingeniería de Telecomunicación, Department of Signal Theory and Communications, University of Vigo, Vigo E-36310, Spain}
\newcommand{\affatlantic}{atlanTTic Research Center, University of Vigo, Vigo E-36310, Spain}

\newcommand{\afftoyama}{Faculty of Engineering, University of Toyama, Gofuku 3190, Toyama 930-8555, Japan}

\begin{document}
\author{Xoel Sixto}
    \email{xsixto@vqcc.uvigo.es}
	\affiliation{\affvqcc} \affiliation{\affuvigo} \affiliation{\affatlantic} 
\author{Álvaro Navarrete}
        \email{anavarrete@vqcc.uvigo.es}
	 	\affiliation{\affvqcc} \affiliation{\affuvigo} \affiliation{\affatlantic} 
\author{Margarida Pereira}
        \affiliation{\affvqcc} \affiliation{\affuvigo} \affiliation{\affatlantic} 
\author{Guillermo Currás-Lorenzo}
        \affiliation{\affvqcc} \affiliation{\affuvigo} \affiliation{\affatlantic} 
\author{Kiyoshi Tamaki}
    \affiliation{\afftoyama} 
\author{Marcos Curty}
\affiliation{\affvqcc} \affiliation{\affuvigo} \affiliation{\affatlantic}

\title{Quantum key distribution with imperfectly isolated devices}
\begin{abstract}
Most security proofs of quantum key distribution (QKD) assume that there is no unwanted information leakage about the state preparation process. However, this assumption is impossible to guarantee in practice, as QKD systems can leak information to the channel due to device imperfections or the active action of an eavesdropper. Here, we solve this pressing issue by introducing a security proof in the presence of information leakage from all state preparation settings for arguably the most popular QKD scheme, namely the decoy-state BB84 protocol. The proof requires minimal experimental characterization, as only a single parameter related to the isolation of the source needs to be determined, thus providing a clear path for bridging the gap between theory and practice. Moreover, if information about the state of the side channels is available, this can be readily incorporated into the analysis to further improve the resulting performance.

\end{abstract}
\maketitle
\section{Introduction}\label{Introduction}

Quantum key distribution (QKD)~\cite{qkd1,qkd2,qkd3} stands as one of the most prominent applications of quantum information science, enabling the establishment of information-theoretic secure communications between two distant parties, often referred to as Alice and Bob. Unlike classical public-key cryptography, whose security relies on computational assumptions, the security of QKD relies on the laws of quantum physics, thus providing protection against any potential eavesdropper ---often referred to as Eve--- with unlimited computational power.

Nevertheless, several challenges still need to be overcome for QKD to become widely adopted. Particularly, it is of paramount importance to develop security proof techniques that can accommodate device imperfections of real-world QKD systems, as any slight deviation between theory and implementation might open a security loophole that could be exploited by Eve \cite{zapatero2024implementation, BSI}. For instance, some security proofs presume that Alice's source generates perfect single-photon states, which is infeasible with current technology. Instead, practical QKD setups typically employ laser sources emitting phase-randomized weak coherent pulses (PRWCPs). It is well-known that this discrepancy between the security proof and the actual implementation allows Eve to launch a photon-number splitting (PNS) attack \cite{pns3,pns2}, which severely limits the achievable performance of QKD.

To protect against PNS attacks, one of the most successful solutions is undoubtedly the decoy-state method \cite{high_loss,decoy, decoy2}, which enables the use of laser sources while providing a secret-key rate that scales linearly with the channel transmittance \cite{concise}, just like ideal single-photon sources. This technique has been successfully implemented in many QKD experiments \cite{exp_decoy1,exp_decoy2,exp_decoy3,exp_decoy4,exp_decoy5,exp_decoy5b,exp_decoy6,exp_decoy7,exp_decoy8}, including satellite links \cite{sat1,sat2,sat3,sat4} as well as commercial setups \cite{idquantique, toshiba, telecom, qteck, thinkq}. What is more, decoy states are a key ingredient of interference-based QKD schemes with laser pulses~\cite{PT}, like {\it e.g.}, measurement-device-independent (MDI) QKD \cite{MDI} or twin-field (TF) QKD \cite{twin}. These later protocols guarantee security with untrusted receivers, thus offering immunity against the Achilles' heel of QKD implementations, detector control attacks~\cite{BSI}, while simultaneously enabling an extension of the transmission distance.

Nonetheless, the generation of decoy states is not free from vulnerabilities and neither is the encoding of bit-and-basis information into these states \cite{loss_tolerant}. Indeed, in practice, it is very hard (if not impossible) to perfectly isolate the users' devices, an issue that is not exclusive from decoy-state implementations but common to all QKD setups \cite{pereira_2019, GLLP, Guille_framework}. This means that information about the state preparation process can be leaked to the channel. This could occur through several mechanisms, like, {\it e.g.}, acoustic or electromagnetic radiation~\cite{baliuka2023deep}, power consumption, or as response to external signals injected by Eve in a Trojan-horse attack (THA)~\cite{vakhitov2001,gissin}. Recently, the security of this latter scenario has been partially addressed in~\cite{tha3, tha2, tha1}. Precisely, these works consider information leakage from a subset of the state preparation settings, particularly from the bit, basis, and intensity values. However, they disregard information leakage from the global phase of the PRWCPs, thus falling short in guaranteeing the security of actual implementations. In addition \cite{tha3, tha2} assume that the quantum state of the side channel is fully characterized, which might be overoptimistic, as it may belong to an infinite-dimensional space and, in general, it may be difficult to measure experimentally.

In this paper, we address these pressing issues by introducing a security proof for decoy-state QKD with imperfectly isolated devices. Importantly, it incorporates information leakage from {\it all} state preparation settings, thus providing a clear path for bridging the gap between theory and practice. Moreover, the proof does not require to experimentally characterize the state of the side channels, but only an upper bound on a parameter that quantifies the deviation of the leaky transmitted states with respect to some predefined ideal states. This fact significantly simplifies its applicability with respect to previous proofs, where the side-channel states needed to be known precisely. In addition, it is valid both for QKD setups that use either passive or active global phase-randomization. Crucially as well, if further information about the side channels is known, this information can be readily incorporated into the security analysis to improve the resulting secret key rate. We illustrate this latter point by proving the security of decoy-state QKD against some THAs similar to those evaluated in~\cite{tha3, tha2, tha1}, in which Eve's back-reflected light is assumed to be a coherent state. In contrast to~\cite{tha3, tha2, tha1}, however, now such coherent state also leaks information about the global phase of Alice's emitted pulses. Finally, as a side contribution, we use the tools developed to quantify the effectiveness of using an external phase modulator (for additional phase randomization) as a countermeasure to THAs. This countermeasure has been suggested in various previous works (see {\it e.g.,}~\cite{gissin,tha2}) but its quantitative analysis has proved elusive until now. 

The paper is organized as follows. In \cref{sec:protocol}, we introduce the decoy-state QKD protocol considered, and outline the main assumptions of the security analysis. \cref{sec:security} contains the main result of this work. Here, we present a security proof that can incorporate information leakage from all state preparation settings and, at the same time, requires minimum characterization of the state of the side channel. \cref{sec:sim} includes simulation results about the expected secret key rates as a function of the magnitude of the side channel. Next, in~\cref{sec:practical}, we address the scenario in which the side channel is assumed to be characterized, and show how this information can be incorporated in the security analysis to improve the system performance. For this, we consider a THA in which the state of Eve's back-reflected light is a coherent state. In this section we also quantitatively evaluate the usefulness of using an external phase modulator to protect against this type of attack. Finally, \cref{sec:conclusions} concludes the paper with a summary of the principal findings. To improve readability, some technical derivations have been moved to a series of Appendices rather than being included in the main text.

\section{Protocol description and transmitted states}\label{sec:protocol}

{\it Protocol description.---} In each round of the protocol, Alice (1) selects an intensity setting $\beta\in\{\mu,\nu,\omega\}$ with probability $p_{\beta}$, (2) selects a certain bit/basis encoding $a\in\{0_Z,1_Z,0_X,1_X\}$ with probability $p_a$ and (3) sends Bob an optical pulse encoded with the chosen settings, and whose global phase $\theta$ follows a probability density function (PDF) $g(\theta)$.

On the receiver side, for each incoming signal, Bob (4) selects a basis $b\in\{Z,X\}$ with probability $p_b$  and (5) performs a measurement described by a positive operator-valued measure (POVM). After that, (6) Bob announces the detected rounds, and for these rounds, both Alice and Bob disclose their basis choices, while Alice also reveals her intensity settings. Finally (7), Alice and Bob extract the secret key from the $Z$-basis events with the signal intensity $\mu$, while they use the other events for parameter estimation.

{\it Assumptions.---} For simplicity, throughout the paper, we shall consider that $(a)$ $p_{0_\alpha}=p_{1_\alpha}$ for $\alpha\in\{Z,X\}$. Also, we shall assume $(b)$ the standard basis independent detection efficiency condition, which implies that the probability that Bob obtains a conclusive measurement outcome for any input state does not depend on his measurement basis choice. Note that this assumption could be relaxed by applying the ideas in \cite{marga_sug} or removed by using  MDI-QKD \cite{MDI} or TF-QKD \cite{twin}. Furthermore, we assume for simplicity that $(c)$ the source generates uncorrelated pulses between rounds and that $(d)$ there are no correlations between the different settings in a given round.

Importantly, we do not assume any specific encoding scheme, as the analysis presented below is valid for any specific encoding method, including \textit{e.g.}, polarization, phase, or time-bin encoding. Throughout this work, we consider the following scenario where the state of a transmitted optical pulse is expressed as below. 

{\it Transmitted states.---}Specifically, let $\hat{V}_B^a$ be an isometric operation for encoding the bit/basis information $a$, performed by the encoder module in Alice's lab, and $\ket{\rm v}_E$ a non-informative quantum state ---say, $e.g.$, the vacuum state--- of a potential side-channel system $E$. The states prepared by Alice in the absence of information leakage can be simply written as
\begin{equation}
\begin{aligned}
\label{eq:ideal}
\rho_{g(\theta),BE}^{\beta,a}&=\int_{0}^{2\pi} g(\theta)\hat{P}\left[\hat{V}_B^a\ket*{\sqrt{\beta}e^{i\theta}}_B\otimes\ket{\rm v}_E\right] d\theta\\
&=\sum_{n}{p}_{n|\beta}\PP{\hat{V}_B^a\ket{n_{\beta, g(\theta)}}_B}\otimes\dyad{\rm v}_E,
\end{aligned}
\end{equation}
where $\hat{P}\left[\ket{\phi}\right]=\dyad{\phi}{\phi}$, the state $\ket*{\sqrt{\beta}e^{i\theta}}_B$ is a coherent state of intensity $\beta$ and global phase $\theta$, the states $\hat{V}_B^a\ket{n_{\beta, g(\theta)}}_B\ket{\rm v}_E$ are the eigenvectors of $\rho_{g(\theta),BE}^{\beta,a}$, and their associated eigenvalues satisfy $\sum_n p_{n|\beta}=1$. Note that here we have eliminated the dependence on $g(\theta)$ of the probabilities ${p}_{n|\beta}$ for simplicity of notation.

In practice, however, due to imperfections of the transmitter's components, or due to Eve's intervention, the actual states prepared by Alice might deviate from those described by \cref{eq:ideal}. In particular, for given values of the parameters $\beta$, $a$ and $\theta$, the actual emitted states can always be written as \cite{pereira_2019, Guille_framework}
\begin{equation}\label{eq:real_states}
\ket{\varphi^{\epsilon}_{\beta,a,\theta}}_{BE}=\sqrt{1-\epsilon}\ket{\varphi_{\beta,a,\theta}}_{BE}+\sqrt{\epsilon}\ket{\varphi^{\perp}_{\beta,a, \theta}}_{BE},
\end{equation}
for some $\epsilon\geq0$, where $\ket{\varphi_{\beta,a, \theta}}_{BE}=\hat{V}_B^a\ket{\sqrt{\beta}e^{i\theta}}_B\otimes\ket{\rm v}_E$ is the desired state and  $\ket*{\varphi^{\perp}_{\beta,a,\theta}}_{BE}$ is a state orthogonal to $\ket*{\varphi_{\beta,a, \theta}}_{BE}$. That is, \cref{eq:real_states} simply expresses the arbitrary state $\ket*{\varphi^{\epsilon}_{\beta,a, \theta}}_{BE}$ in a certain orthonormal basis. Moreover, since, in practice, it might be very difficult to determine the side-channel state $\ket*{\varphi^{\perp}_{\beta,a,\theta}}_{BE}$, for now we shall consider that this state is completely uncharacterized.  For instance, in a THA we have that
\begin{equation}
\label{eq:THA_generalstate}
\ket*{\varphi^{\epsilon}_{\beta,a,\theta}}_{BE}=\hat{V}_B^a\ket*{\sqrt{\beta}e^{i\theta}}_B\ket{\lambda_{\beta,a,\theta}}_E,
\end{equation}
where the state of the back-reflected light from Eve has the form $\ket*{\lambda_{\beta,a,\theta}}_E=\sqrt{1-\epsilon}\ket{\rm v}_E+\sqrt{\epsilon}\ket*{u_{\beta,a,\theta}}_E$ with $\ket*{u_{\beta,a,\theta}}_E$ being orthogonal to the vacuum state and in general uncharacterized in a conservative scenario. That is, in this particular case, we find that 
\begin{equation}
\label{eq:dos}
\ket*{\varphi^{\perp}_{\beta,a, \theta}}_{BE}=\hat{V}_B^a\ket*{\sqrt{\beta}e^{i\theta}}_B\ket*{u_{\beta,a,\theta}}_E.    
\end{equation}

Putting it all together, we have that in the presence of general information leakage about the settings $\beta$, $a$ and $\theta$, the actual states emitted by Alice are given by
\begin{equation}
\label{eq:states}
\begin{aligned}
\rho_{g(\theta),BE}^{\epsilon,\beta,a}&=\int_{0}^{2\pi}g(\theta) \PP{\ket{\varphi^{\epsilon}_{\beta, a, \theta}}_{BE}}d\theta\\    &=\sum_{n}p^{\epsilon}_{n|\beta,a}\PP{\ket*{n^{\epsilon}_{\beta, a, g(\theta)}}_{BE}},
\end{aligned}
\end{equation}
where now the eigenvectors $\ket*{n^{\epsilon}_{\beta, a, g(\theta)}}_{BE}$ and the eigenvalues $p^{\epsilon}_{n|\beta,a}$  are unknown because the states $\ket*{\varphi^{\perp}_{\beta,a, \theta}}_{BE}$ are generally uncharacterized. Still, when the parameter $\epsilon$ is very small one expects that $\ket*{n^{\epsilon}_{\beta, a, g(\theta)}}_{BE}\approx \hat{V}_B^a\ket{n_{\beta, g(\theta)}}_{B}\otimes\ket{\rm v}_E$ and $p^{\epsilon}_{n|\beta,a}\approx{p}_{n|\beta}$.

\subsection{Passive versus active global phase randomization}

\begin{figure}[htp]
\centering
\begin{minipage}{\columnwidth}
\centering
\includegraphics[width=\columnwidth]{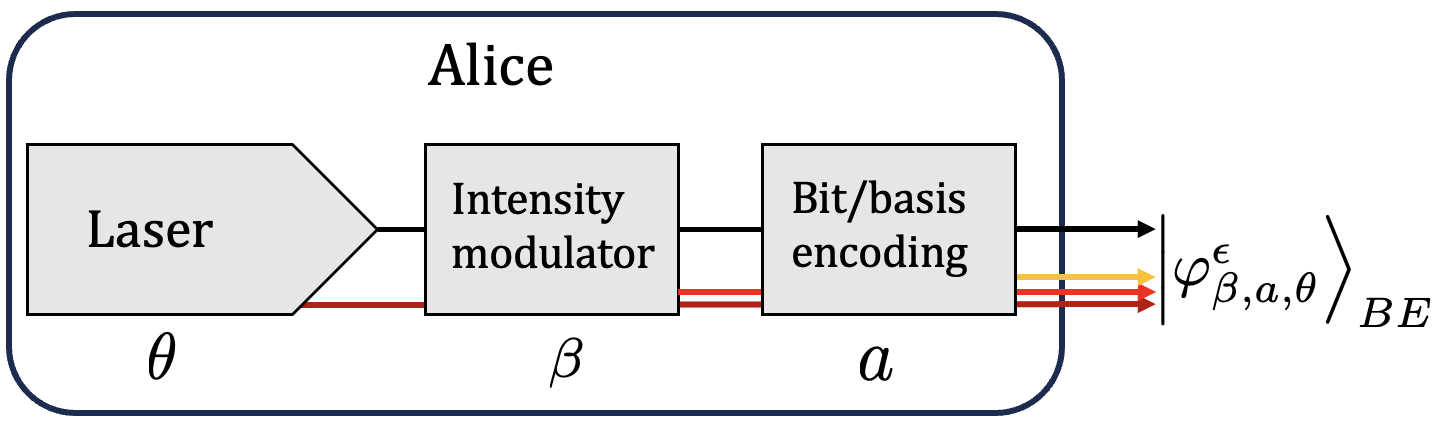}
(a)
\end{minipage}
\bigskip
\begin{minipage}{\columnwidth}
\centering
\includegraphics[width=\columnwidth]{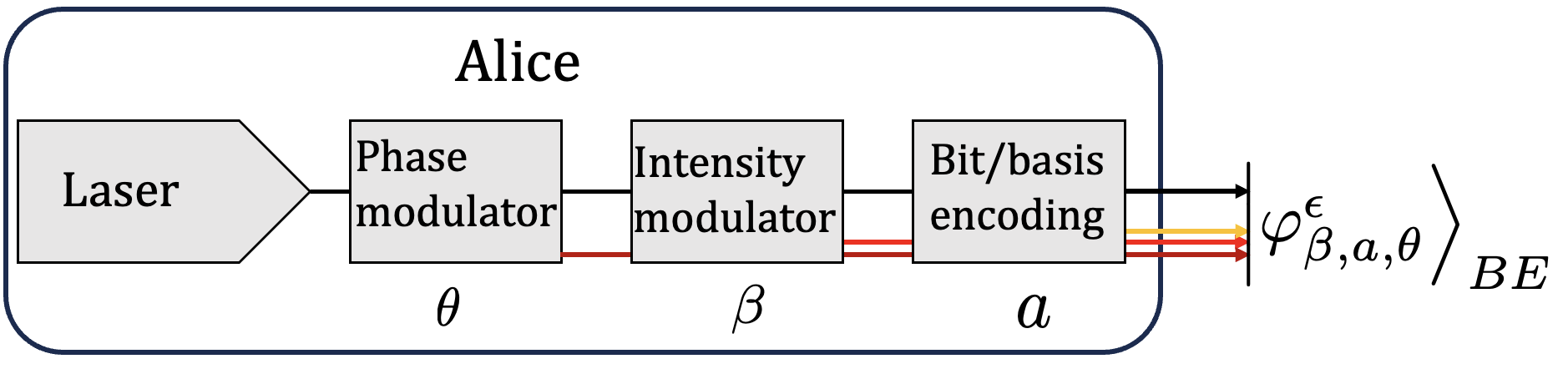}
(b)
\end{minipage}
\caption{\label{fig:figura_principal}Two examples of decoy-state BB84 transmitters leaking information about the global phase $\theta$, the intensity $\beta$, and the bit/basis value $a$ to the quantum channel. In (a) the global phase $\theta$ is randomized passively ---$e.g.$, by gain-switching the laser--- while in (b) the phase randomization is achieved actively by employing a phase modulator. The state $\ket{\varphi^{\epsilon}_{\beta,a,\theta}}_{BE}$ is given by \cref{eq:real_states} and describes the signal that Alice sends to Bob together with the side channel that leaks information to Eve for given values of $\beta$, $a$ and $\theta$.}
\end{figure}

We note that the security proof we introduce later in \cref{sec:security} can be applied to the two practical scenarios shown in \cref{fig:figura_principal}. The first one pertains to passive phase randomization, in which the global phase of each emitted pulse is randomized within the laser source via gain-switching (see \cref{fig:figura_principal}a). In the absence of imperfections, this results in a uniform phase distribution, $i.e.$, $g(\theta)\equiv f(\theta)=(2\pi)^{-1}$. This means that in an ideal scenario, the statistics ${p}_{n|\beta}$ in \cref{eq:ideal} follow a Poisson distribution, and the signals $\ket{n_{\beta,f(\theta)}}_B$ are Fock states with $n$ photons. When information about the state preparation process is leaked to the channel, on the other hand, the emitted states can be directly obtained from \cref{eq:states} by particularizing to this $g(\theta)$. 

The second scenario involves active phase randomization, which can be achieved, $e.g.$, by placing an external phase modulator at the output of the laser source (see \cref{fig:figura_principal}b). In particular, if we consider that Alice selects the phase of each pulse within a set of $N$ possible phases evenly distributed in the interval $[0,2\pi)$, the PDF $g(\theta)$ can be written as
\begin{equation}
\label{eq:pdf}
g(\theta)\equiv h(\theta)=\frac{1}{N}\sum_{k=0}^{N-1}\delta(\theta-\theta_k),
\end{equation}
where $\theta_k=2\pi k /N$.
Thus, in the absence of information leakage, it turns out that the signals $\ket{n_{\beta, h(\theta)}}_B$ in \cref{eq:ideal} are given by \cite{phase0}
\begin{equation}
    \left|n_{\beta, h(\theta)}\right\rangle_{B}=\tilde{N}\sum_{l=0}^{\infty} \frac{\sqrt{\beta}^{~l N+n}}{\sqrt{(l N+n) !}}|l N+n\rangle_B,
\end{equation}
where $\tilde{N}$ represents a normalization factor and we have replaced $g(\theta)$ with the PDF $h(\theta)$ given by \cref{eq:pdf}. Moreover, note that, in this particular case, the sum over $n$ in \cref{eq:ideal} runs from zero to $N-1$. Like in the previous scenario, when some information about the selected internal state preparation settings is leaked to the channel, the actual transmitted states can be directly obtained from \cref{eq:states} by particularizing $g(\theta)$ to the PDF $h(\theta)$ given by \cref{eq:pdf}.
%

Importantly, we remark that our analysis below can be readily extended to accommodate any $g(\theta)$ by using the techniques presented in \cite{phase2}.

\section{Security proof} \label{sec:security}

In this section, we prove the security of decoy-state QKD in the asymptotic regime, {\it i.e.,} in the limit of a large number of pulses sent. Therefore, we neglect the statistical fluctuations due to finite-size effects. Moreover, we shall consider that  Alice's emitted states are given by \cref{eq:states} and the signals $\ket*{\varphi^{\perp}_{\beta,a, \theta}}_{BE}$ are uncharacterized.

In the security proof, we consider an equivalent entanglement-based protocol in which Alice substitutes the rounds where she sends the states $\ket*{n^{\epsilon}_{\beta,0_\alpha, g(\theta)}}_{BE}$ and $\ket*{n^{\epsilon}_{\beta,1_\alpha, g(\theta)}}_{BE}$ by the preparation of an entangled state $\ket*{\Psi^{n,\epsilon}_{\beta, \alpha, g(\theta)}}_{ABE}$ of the form
\begin{equation}
\label{eq:bas_dep_states}
\begin{aligned}
\ket*{\Psi^{n,\epsilon}_{\beta, \alpha, g(\theta)}}_{ABE}=&\sqrt{q_{\beta,0_\alpha}^{n, \epsilon}}\ket{0_\alpha}_A\ket*{n^{\epsilon}_{\beta,0_\alpha, g(\theta)}}_{BE}\\
&+e^{i\phi_\alpha}\sqrt{q_{\beta, 1_\alpha}^{n, \epsilon}}\ket{1_\alpha}_A\ket*{n^{\epsilon}_{\beta,1_\alpha, g(\theta)}}_{BE},
\end{aligned}
\end{equation}
with $\alpha \in \{Z,X\}$ and where $\phi_{\alpha}$ represents an arbitrary phase and
\begin{equation}
\label{eq:qvirtualstate}
\begin{aligned}
q_{\beta, j_\alpha}^{n, \epsilon}=\frac{p^{n, \epsilon}_{\beta, j_\alpha}}{p^{n, \epsilon}_{\beta,0_\alpha}+p^{n,\epsilon}_{\beta,1_\alpha}},
\end{aligned}
\end{equation}
where $p^{n, \epsilon}_{\beta, j_\alpha}=p_{j_\alpha}p_{\beta}p^{\epsilon}_{n|\beta,j_\alpha}$ and $j\in\set{0,1}$. Note that if Alice measures her ancilla system $A$ in the $Z$ basis, we recover the actual scenario.

Now, we define $Q_{\beta, j_{\alpha}}$ ($E_{\beta, j_{\alpha}}$) as the number of the events divided by the number of pulses sent for those rounds in which Alice selects the intensity $\beta$ and the bit value $j$, and both parties choose the basis $\alpha$, $i.e.,$ the so-called overall gain (quantum bit-error rate). Also, let the yield $Y_{\beta, j_{\alpha}}^{n}$ denote the probability that Bob observes a conclusive detection event in his measurement apparatus conditioned on Alice emitting the state $\ket*{n^{\epsilon}_{\beta,j_{\alpha}, g(\theta)}}_{BE}$ and Bob selecting the basis $\alpha$. Note that, in the notation of the parameters $Q_{\beta, j_{\alpha}}$, $E_{\beta, j_{\alpha}}$ and $Y_{\beta, j_{\alpha}}^{n}$ we omit their dependence on $\epsilon$ and $g(\theta)$ for simplicity. Moreover, let 
\begin{equation}
    \begin{aligned}
        &Q_{\beta,  \alpha}:=\frac{1}{2}\sum_{j=0}^{1}Q_{\beta, j_{\alpha}}\\
        &E_{\beta, \alpha}:=\frac{1}{2Q_{\beta,  \alpha}}\sum_{j=0}^{1}Q_{\beta, j_{\alpha}}E_{\beta,j_{\alpha}},\\
        &Y^{n}_{\beta, \alpha}:=\sum_{j=0}^{1}q^{n,\epsilon}_{\beta,j_{\alpha}}Y^{n}_{\beta,j_\alpha},
    \end{aligned}
\end{equation}
represent, respectively, the gain, quantum bit-error rate, and $n$-photon yield corresponding to the intensity $\beta$ and basis $\alpha$. Finally, let $e^{{\rm ph}, n}_{\beta}$ refer to the $n$-photon phase-error rate of the virtual entanglement-based protocol, \textit{i.e.}, the probability that Alice and Bob obtain different outcomes if, in those rounds in which Alice prepares the entangled state $\ket*{\Psi^{n,\epsilon}_{\beta, Z, g(\theta)}}_{ABE}$ given by \cref{eq:bas_dep_states}, they measure their corresponding systems $A$ and $B$ in the $X$ basis.

It can be shown that the asymptotic secret-key rate of the protocol satisfies \cite{decoy, GLLP} 
\begin{equation}\label{eq:skr_general}
\begin{aligned}
R&\geq p_Z^2p_{\mu}\sum_{n} p^{\epsilon}_{n| \mu, Z}Y_{\mu, Z}^{n}\left[1-h\left(e^{{{\rm ph}, n}}_{\mu}\right)\right]-\lambda_{\rm EC} \\
&\geq p_Z^2p_{\mu} p_{1| \mu, Z}^{\epsilon,\text{L}}Y_{\mu, Z}^{1, \text{L}}\left[1-h\left(e_{\mu}^{{\rm ph},1, \mathrm{U}}\right)\right]-\lambda_{\rm EC},
\end{aligned}
\end{equation}
where $p^{\epsilon}_{n| \mu, Z}=\frac{1}{p_Z}(p_{0_Z}p^{\epsilon}_{n| \mu, 0_Z}+p_{1_Z}p^{\epsilon}_{n| \mu, 1_Z})$ with $p_Z=p_{0_Z}+p_{1_Z}$, the function $h(x)=-x\log_2{(x)}-(1-x)\log_2{(1-x)}$ is the binary Shannon entropy function, the parameter $\lambda_{\rm EC}=p_Z^2p_{\mu} f Q_{\mu, Z} h\left(E_{\mu, Z}\right)$ is the number of bits revealed in the error correction phase of the protocol per transmitted signal, $f$ is the efficiency of the error correction scheme, and the superscript L (U) indicates that the quantity is a lower (upper) bound. Note that in the second inequality of~\cref{eq:skr_general} we simply disregard the secret key that Alice and Bob could distill from those states with $n\neq 1$. That is, we assume that they extract key only from the states $\ket*{1^{\epsilon}_{\mu, j_Z, g(\theta)}}_{BE}$ with $j\in\{0,1\}$, although in what follows, we keep the dependence on $n$ for generality.

In the next subsection, we show how to estimate both a lower bound on the yield $Y_{\mu, Z}^{1}$ and an upper bound on the phase-error rate $e_{\mu}^{{\rm ph},1}$ in the presence of information leakage.

\subsection{Yield estimation}

In the standard decoy-state analysis, a fundamental premise is that $n$-photon states are independent of the intensity setting and, as a result of this, their detection statistics (or so-called yields) associated to different intensity settings are the same. However, when the $n$-photon states depend on the intensity setting, this condition no longer holds as Eve can behave differently for different intensity settings. To put it in other words, the statistical outcomes of Eve's operation on the $n$-photon states become different for different intensity settings.

To relate the yields associated with different intensities, we shall invoke the quantum-coin argument \cite{GLLP, lo_preskill,phase0, Guille_framework} (see Appendix~\ref{app:CS}). Importantly as shown in \cite{pereira_2020}, this allows us to impose some constraints ---which, in line with~\cite{Zapatero_2021}, we shall refer to as the Cauchy-Schwarz (CS) constraints--- to establish the following bounds between yields with different intensity settings
\begin{equation}
\label{cs_text}
\begin{aligned}
&G_{-}\left(Y^{n}_{\zeta, a},\vert\braket*{n_{\zeta,a, g(\theta)}^{\epsilon}}{n_{\gamma,a, g(\theta)}^{\epsilon}}_{BE}\vert^2 \right) \leq Y^{n}_{\gamma, a} \\&\leq 
 G_{+}\left(Y^{n}_{\zeta, a},\vert\braket*{n_{\zeta,a, g(\theta)}^{\epsilon}}{n_{\gamma,a, g(\theta)}^{\epsilon}}_{BE}\vert^2 \right),
\end{aligned}
\end{equation}
with $\zeta,\gamma\in\{\mu,\nu,\omega\}$ and where the functions $G_{\pm}$ are defined as
\begin{equation}
\label{eq6}
\begin{aligned}
&G_{-}(y, z)= \begin{cases} g_{-}(y, z) & \text { if } y>1-z, \\
0 & \text { otherwise},\end{cases}\\
&G_{+}(y, z)= \begin{cases}g_{+}(y, z) & \text { if } y<z, \\
1 & \text { otherwise},\end{cases}
\end{aligned}
\end{equation}
with
\begin{equation}
\label{eq:smallg}
g_{\pm}(y, z)=y+(1-z)(1-2 y) \pm 2 \sqrt{z(1-z) y(1-y)}.
\end{equation}

Next, we connect the yields with the observables of the QKD protocol via the following decoy-state constraints
\begin{equation}
\begin{aligned} \label{eq:gain_1}
&Q_{\beta,  a}\geq \sum_{n=0}^{n_{\rm cut}}p^{\epsilon, \text{L}}_{n|\beta, a}Y^{n}_{\beta, a},\\
&Q_{\beta,  a}\leq 1- \sum_{n=0}^{n_{\rm cut}}p^{\epsilon, \text{L}}_{n|\beta, a}(1-Y^{n}_{\beta, a}),
\end{aligned}
\end{equation}
which hold for any $n_{\rm cut}\geq 0$ and where the bound 
\begin{equation}
\label{eq:bound_main}
p^{\epsilon, \text{L}}_{n|\beta, a}= \max\{p_{n|\beta}-\sqrt{\epsilon}, 0\}  
\end{equation}
is introduced in \cref{app:use} with $\sqrt{\epsilon}$ being half of the trace distance between the states in \cref{eq:ideal,eq:states}. This represents the maximum deviation between the probability to emit the $n$-photon state from the perfect state in \cref{eq:ideal} and the one with side-channel in \cref{eq:states}. This is because the statistics $p^{\epsilon}_{n|\beta,a}$ are unknown and thus one must obtain lower bounds on them. Also, we remark that for the discrete phase randomization scenario represented in \cref{fig:figura_principal}b, the sum in $n$ is no longer up to infinity, but up to $N-1$, and so one could directly set $n_{\rm cut}=N-1$. Besides, in that particular scenario, we have that the second constraint in \cref{eq:gain_1} can be replaced by 
\begin{equation}
\label{eq:gain_discrete}
\begin{aligned}
Q_{\beta,  a}\leq \sum_{n=0}^{N-1}p^{\epsilon, \text{U}}_{n|\beta, a}Y^{n}_{\beta,a},
\end{aligned}
\end{equation}
where the bound 
\begin{equation}
p^{\epsilon, \text{U}}_{n|\beta, a}= \min\{p_{n|\beta}+\sqrt{\epsilon}, 1\}
\end{equation} 
is defined in a similar manner like \cref{eq:bound_main} (see \cref{app:use} for further details). According to our simulations, this latter constraint results in a  tighter estimation that than in \cref{eq:gain_1}.

To finish with, we construct a linear program to estimate a lower bound on the desired yield $Y^{1}_{\mu, Z}$. Nevertheless, to enable the use of linear programming, in \cref{app:linear_CS}  we introduce a linearized version of \cref{cs_text}, which requires to define some reference yields $\tilde{Y}^{n}_{\beta, a}$ \cite{Zapatero_2021}. This linearized version still requires to bound the quantities  $\abs*{\braket*{n_{\zeta,a, g(\theta)}^{\epsilon}}{n_{\gamma,a, g(\theta)}^{\epsilon}}_{BE}}^2$ that appear in \cref{cs_text}. This can be done with semidefinite programming (SDP), as shown in \cref{sec:inner_SDP}, by adapting the techniques recently presented in  \cite{Guille_framework} to our study. The resulting linear program is given by
\begin{widetext}
\begin{gather}\label{LP_final}
\begin{aligned} 
\textup{min}\hspace{.1cm}& ~Y^{n}_{\beta, a} \\
\textup{s.t.}\hspace{.1cm}&G_{-}\left(\tilde{Y}^{n}_{\zeta, a}, \vert\braket*{n_{\zeta,a, g(\theta)}^{\epsilon}}{n_{\gamma,a, g(\theta)}^{\epsilon}}_{BE}\vert^2\right)+G_{-}^{\prime}\left(\tilde{Y}^{n}_{\zeta, a}, \vert\braket*{n_{\zeta,a, g(\theta)}^{\epsilon}}{n_{\gamma,a, g(\theta)}^{\epsilon}}_{BE}\vert^2\right)\left(Y^{n}_{\zeta, a}-\tilde{Y}^{n}_{\zeta, a}\right) \leq Y^{n}_{\gamma, a}  \\
&\leq G_{+}\left(\tilde{Y}^{n}_{\zeta, a}, \vert\braket*{n_{\zeta,a, g(\theta)}^{\epsilon}}{n_{\gamma,a, g(\theta)}^{\epsilon}}_{BE}\vert^2\right)+G_{+}^{\prime}\left(\tilde{Y}^{n}_{\zeta, a}, \vert\braket*{n_{\zeta,a, g(\theta)}^{\epsilon}}{n_{\gamma,a, g(\theta)}^{\epsilon}}_{BE}\vert^2\right)\left(Y^{n}_{\zeta, a}-\tilde{Y}^{n}_{\zeta, a}\right),\\
&Q_{\gamma,  a}\geq \sum_{n=0}^{n_{\rm cut}}p^{\epsilon, \text{L}}_{n|\gamma, a}Y^{n}_{\gamma, a},\\
&Q_{\gamma,  a}\leq 1- \sum_{n=0}^{n_{\rm cut}}p^{\epsilon, \text{L}}_{n|\gamma, a}(1-Y^{n}_{\gamma, a}),\\
&0\leq Y^{n}_{\gamma, a} \leq 1,\\
& ~\forall \zeta, \gamma \in \{\mu, \nu, \omega \},\ \zeta\neq\gamma,\\
&~\forall n\in\{0,...,n_{\rm cut}\}.
\end{aligned}
\end{gather}
\end{widetext}

The exact form of the functions $G_{\pm}^{'}$, together with a discussion on the values of the reference parameters $\tilde{Y}^{n}_{\beta, a}$ are included in \cref{app:linear_CS}. Finally, note that for the discrete scenario, one can add to the constraints in \cref{LP_final} the gain constraint given in \cref{eq:gain_discrete}.

By solving \cref{LP_final}, it is possible to compute a lower bound on the yields associated to different encoding settings separately. Then, a lower bound on the $n$-photon $Z$-basis yield is given by
\begin{equation}\label{eq:min}
Y^{n}_{\mu, Z}=\sum_{j=0}^{1}q_{\mu, j_Z}^{n, \epsilon}Y^{n}_{\mu, j_Z}
\geq\min_{j}Y^{n}_{\mu, j_Z}  =: Y^{n,L}_{\mu, Z}.
\end{equation}

\subsection{Phase-error rate estimation}\label{sec:phase_error}

Having established the equivalence between the actual protocol and the emission of a quantum coin, as shown in \cref{app:coin}, one can upper bound the $n$-photon phase-error rate,  $e^{{\rm ph}, n}_{\mu}$ from an upper bound $e_{\mu, X}^{{\rm b}, n, \rm{U}}$ on the $X$-basis bit-error rate $e_{\mu, X}^{{\rm b}, n}$ as \cite{GLLP}
\begin{equation}\label{qwe}
\begin{aligned}
&e^{{\rm ph}, n, {\rm U}}_{\mu} \leq e_{\mu, X}^{{\rm b}, n, \rm{U}}+
4 \Delta^{n}_{\mu}\left(1-\Delta^{n}_{\mu}\right)\left(1-2 e_{\mu, X}^{{\rm b}, n, \rm{U}}\right)\\
&+4\left(1-2 \Delta^{n}_{\mu}\right) \sqrt{\Delta^{n}_{\mu}\left(1-\Delta^{n}_{\mu}\right) e_{\mu, X}^{{\rm b}, n, \rm{U}}\left(1-e_{\mu, X}^{{\rm b}, n, \rm{U}}\right)},
\end{aligned}
\end{equation}
where
\begin{equation}
\label{eq:basis_dependence}
\Delta^{n}_{\beta}=\frac{1-\sqrt{F^{n,\epsilon,\rm{L}}_{\beta}}}{2 Y^{n,\rm{L}}_{\beta, \rm{coin}}}, 
\end{equation}
is an upper bound on the quantum-coin imbalance of the detected coin rounds, the parameter 
\begin{equation}
Y^{n,\rm{L}}_{\beta, \rm{coin}}= \min_{\alpha}Y^{n, \rm{L}}_{\beta, \alpha},
\end{equation}
is a lower bound on the yield of the quantum coin, $Y^{n,\rm{L}}_{\beta, X}= \min_{j}Y^{n}_{\beta, j_X}$ can be computed using~\cref{LP_final,eq:min}, $Y^{n,{\rm L}}_{\beta,Z}$ is given by \cref{eq:min} and $F^{n,\epsilon,\rm{L}}_{\beta}$ is a lower bound on the fidelity 
\begin{equation}
F^{n,\epsilon}_{\beta}:=F\big(|\Psi^{n, \epsilon}_{\beta, Z, g(\theta)}\rangle_{ABE},|\Psi^{n, \epsilon}_{\beta, X, g(\theta)}\rangle_{ABE}\big).
\end{equation}
 
To obtain $F^{n,\epsilon,\rm{L}}_{\beta}$, we take advantage of the Bures distance, which is defined as \cite{bures}
\begin{equation}
\begin{aligned}
\label{eq:bures}
&d_B(\ket*{\Psi^{n, \epsilon}_{\beta, Z, g(\theta)}}_{ABE},\ket*{\Psi^{n, \epsilon}_{\beta, X, g(\theta)}}_{ABE})^2\\
&=2-2\sqrt{F\left(\ket*{\Psi^{n, \epsilon}_{\beta, Z, g(\theta)}}_{ABE}, \ket*{\Psi^{n, \epsilon}_{\beta, X, g(\theta)}}_{ABE}\right)}.
\end{aligned}
\end{equation}
Importantly, this quantity obeys the triangle inequality and consequently, we have that
\begin{equation}
\label{eq:triangle}
\begin{aligned}
&d_B\left(\ket*{\Psi^{n,\epsilon}_{\beta, Z, g(\theta)}}_{ABE}, \ket*{\Psi^{n,\epsilon}_{\beta, X, g(\theta)}}_{ABE}\right) \\
&\leq d_B\left(\ket*{\Psi^{n,\epsilon}_{\beta, Z, g(\theta)}}_{ABE}, \ket*{\Psi^{n,{\rm ideal}}_{\beta, Z, g(\theta)}}_{ABE}\right)\\
&+d_B\left(\ket*{\Psi^{n,{\rm ideal}}_{\beta, Z, g(\theta)}}_{ABE}, \ket*{\Psi^{n,{\rm ideal}}_{\beta, X, g(\theta)}}_{ABE}\right)\\
&+ d_B\left(\ket*{\Psi^{n,{\rm ideal}}_{\beta, X, g(\theta)}}_{ABE}, \ket*{\Psi^{n,\epsilon}_{\beta, X, g(\theta)}}_{ABE}\right),
\end{aligned}
\end{equation}
where
\begin{equation}
\label{eq:bas_dep_states_ideal}
\begin{aligned}
\ket*{\Psi^{n, {\rm ideal}}_{\beta, \alpha, g(\theta)}}_{ABE}=\frac{1}{\sqrt{2}} \sum_{j=0}^{1}\ket*{j_{\alpha}}\ket*{n_{\beta,j_\alpha, g(\theta)}}_{BE}
\end{aligned}
\end{equation}
represents the ideal version of the states introduced in \cref{eq:bas_dep_states} in the absence of information leakage, $i.e.,$ $\ket*{\Psi^{n, {\rm ideal}}_{\beta, \alpha, g(\theta)}}_{ABE}\equiv\ket*{\Psi^{n,\epsilon=0}_{\beta, \alpha, g(\theta)}}_{ABE}$, and where we have set $\phi_\alpha=0$ for convenience.

In doing so, we divide the problem of estimating a lower bound on the fidelity $F^{n,\epsilon}_{\beta}$ in three separate steps given by \cref{eq:triangle}. Finally, $F^{n,\epsilon,\rm{L}}_{\beta}$ can be obtained straightforwardly from an upper bound on $d_B(\ket*{\Psi^{n,\epsilon}_{\beta, Z, g(\theta)}}_{ABE}, \ket*{\Psi^{n,\epsilon}_{\beta, X, g(\theta)}}_{ABE})$ by means of \cref{eq:bures}.

The Bures distance $d_B(\ket*{\Psi^{n,{\rm ideal}}_{\beta, Z, g(\theta)}}_{ABE}, \ket*{\Psi^{n,{\rm ideal}}_{\beta, X, g(\theta)}}_{ABE})$, can be easily computed, as the ideal states are completely characterized given $g(\theta)$. For instance, if the PDF $g(\theta)$ is given by $g(\theta)=f(\theta)=(2\pi)^{-1}$ ($i.e.$, in the case of perfect passive phase randomization) we have that $\ket{n_{\beta,j_\alpha,f(\theta)}}_{B}=\hat{V}_{B}^{j_\alpha}\ket{n}_B$ for $j\in\{0,1\}$, and so, $d_B(\ket*{\Psi^{n,{\rm ideal}}_{\beta, Z, f(\theta)}}_{ABE}, \ket*{\Psi^{n,{\rm ideal}}_{\beta, X, f(\theta)}}_{ABE})=0$ for $n=1$. Similar calculations can be made for any other photon number $n$ and phase distribution function $g(\theta)$ applying the techniques in \cite{phase2}. 

Next, we show how to upper bound the Bures distance $d_B(\ket*{\Psi^{n,\epsilon}_{\beta, \alpha, g(\theta)}}_{ABE}, \ket*{\Psi^{n, {\rm ideal}}_{\beta, \alpha, g(\theta)}}_{ABE})$ (or, equivalently, the fidelity) between the real and the ideal states as defined in \cref{eq:bas_dep_states,eq:bas_dep_states_ideal} respectively. For this, we note that
\begin{equation}
\begin{aligned}
\label{eq:fide_bound_2}
F&(\ket*{\Psi^{n,\epsilon}_{\beta,\alpha, g(\theta)}},\ket*{\Psi^{n,\rm{ideal}}_{\beta,\alpha, g(\theta)}})=|\langle \Psi^{n,\epsilon}_{\beta,\alpha, g(\theta)} |\Psi^{n,\rm{ideal}}_{\beta,\alpha, g(\theta)}\rangle|^2\\
=&\Bigg|\sqrt{\frac{q_{\beta,0_\alpha }^{n, \epsilon}}{2}}\langle n^{\epsilon}_{\beta,0_\alpha, g(\theta)}|n_{\beta,0_\alpha, g(\theta)} \rangle \\
&+e^{-i\phi_{\alpha}}\sqrt{\frac{q_{\beta, 1_\alpha}^{n, \epsilon}}{2}}\langle n^{\epsilon}_{\beta,1_\alpha, g(\theta)} | n_{\beta,1_\alpha, g(\theta)}\rangle\Bigg|^2 \\
=&\Bigg(\sum_{j=0}^{1}\sqrt{\frac{q_{\beta, j_\alpha}^{n,\epsilon}}{2}}|\langle n^{\epsilon}_{\beta,j_\alpha, g(\theta)}|n_{\beta,j_\alpha, g(\theta)} \rangle|\Bigg)^2,
\end{aligned}
\end{equation}
where we have omitted the subscripts $ABE$ in the states for simplicity. Note that in \cref{eq:fide_bound_2} the last equality always holds for certain $\phi_{\alpha}$. Importantly, both the probabilities $q_{\beta, j_\alpha}^{n, \epsilon}$ and the inner products $|\langle n^{\epsilon}_{\beta,j_\alpha, g(\theta)}|n_{\beta,j_\alpha, g(\theta)} \rangle|$ that appear in this equation can be bounded using the results presented in \cref{app:use}. 

Putting it all together, and proceeding in a similar way to the yield calculation of the previous section, we find that an upper bound on the bit-error rate $e_{\beta, X}^{{\rm b}, n}$ can be obtained by solving the following linear program
\begin{widetext}
\begin{gather}\label{LP_final_error}
\begin{aligned} 
\textup{max}\hspace{.1cm}& ~\xi^{n}_{\beta, j_X} \\
\textup{s.t.}\hspace{.1cm}&G_{-}\left(\tilde{\xi}^{n}_{\zeta, j_X}, \vert\braket*{n_{\zeta,j_X, g(\theta)}^{\epsilon}}{n_{\gamma,j_X, g(\theta)}^{\epsilon}}_{BE}\vert^2\right)+G_{-}^{\prime}\left(\tilde{\xi}^{n}_{\zeta, j_X}, \vert\braket*{n_{\zeta,j_X, g(\theta)}^{\epsilon}}{n_{\gamma,j_X, g(\theta)}^{\epsilon}}_{BE}\vert^2\right)\left(\xi^{n}_{\zeta, j_X}-\tilde{\xi}^{n}_{\zeta, j_X, g(\theta)}\right) \leq \xi^{n}_{\gamma, j_X} \\
&\leq G_{+}\left(\tilde{\xi}^{n}_{\zeta, j_X}, \vert\braket*{n_{\zeta,j_X,g(\theta)}^{\epsilon}}{n_{\gamma,j_X,g(\theta)}^{\epsilon}}_{BE}\vert^2\right)+G_{+}^{\prime}\left(\tilde{\xi}^{n}_{\zeta, j_X}, \vert\braket*{n_{\zeta,j_X,g(\theta)}^{\epsilon}}{n_{\gamma,j_X,g(\theta)}^{\epsilon}}_{BE}\vert^2\right)\left(\xi^{n}_{\zeta, j_X}-\tilde{\xi}^{n}_{\zeta, j_X}\right),\\
&E_{\gamma,  j_X} Q_{\gamma,  j_X}\geq \sum_{n=0}^{n_{\rm cut}}p^{\epsilon, \text{L}}_{n|\gamma, j_X}\xi^{n}_{\gamma, j_X},\\
&E_{\gamma,  j_X} Q_{\gamma,  j_X}\leq 1- \sum_{n=0}^{n_{\rm cut}}p^{\epsilon, \text{L}}_{n|\gamma, j_X}(1-\xi^{n}_{\gamma, j_X}),\\
&0\leq \xi^{n}_{\gamma, j_X} \leq 1\\
&~\forall \zeta, \gamma \in \{\mu, \nu, \omega \},\ \zeta\neq\gamma,\\
&~\forall n\in\{0,...,n_{cut}\},
\end{aligned}
\end{gather}
\end{widetext}
where  $j\in\{0,1\}$,  $\xi^{n}_{\beta, j_X}=e^{{\rm b}, n}_{\beta, j_X} Y^{n}_{\beta, j_X}$ is the bit-error probability associated with the transmitted states $\ket*{n^{\epsilon}_{\beta, j_X, g(\theta)}}_{BE}$, and $\tilde{\xi}^{n}_{\zeta, j_X}$ is a reference value for $\xi^{n}_{\zeta, j_X}$ whose selection can be optimized for each distance (see \cref{app:linear_CS} for more details). Thus, by using  \cref{LP_final_error} one can compute an upper bound on the $X$-basis bit-error rate by first obtaining
\begin{equation}\label{eq:min_phase}
\xi^{n}_{\beta, X} =\sum_{j=0}^1 q_{\beta, j_X}^{n, \epsilon}\xi^{n}_{\beta, j_X}
\leq\max_{j}\xi^{n}_{\beta, j_X}=:\xi^{n, \rm{U}}_{\beta,X},
\end{equation}
and then calculating 
\begin{equation}\label{eq:phase_error_bound}
e^{{\rm b}, n}_{\beta, X}\leq\frac{\xi^{n,\rm{U}}_{\beta, X}}{Y^{n, {\rm L}}_{\beta, X}}=:e^{{\rm b}, n, {\rm U}}_{\beta,X}.
\end{equation}

From \cref{qwe} and the parameters $e^{{\rm b}, 1, {\rm U}}_{\beta,X}$, $Y^{1,\rm{L}}_{\beta, \rm{coin}}$ and $F^{1,\epsilon,\rm{L}}_{\beta}$, one can determine an upper bound $e^{{\rm ph},1, {\rm U}}_{\mu}$ on the single-photon phase error rate.

Finally, after obtaining $Y^{1, {\rm L}}_{\mu, Z}$ from~\cref{eq:min} and $e^{{\rm ph},1, {\rm U}}_{\mu}$ from~\cref{qwe}, one can evaluate the lower bound on the secret-key rate given by \cref{eq:skr_general}. This is what we do in the next section.

\section{Simulations}\label{sec:sim}

 For the simulations, we use a typical channel model to compute the expected gains $Q_{\beta, \alpha}$ and error rates $E_{\beta, \alpha}$ that Alice and Bob would observe in a real implementation of the scheme (see \cref{app:channel_model}). In addition, we consider that the error correction efficiency is $f=1.16$. Note that, since we focus in the asymptotic secret-key rate scenario, we can assume that $p_{\mu}\approx 1$ and $p_{Z}\approx1$, which maximize the secret-key rate. Finally, we optimize the intensities $\mu$ and $\nu$ for each value of the distance and set the intensity of the vacuum signal for simplicity to $\omega=0$. 

\begin{figure}[H]
\centering\includegraphics [width= 8.6cm, height=6cm] {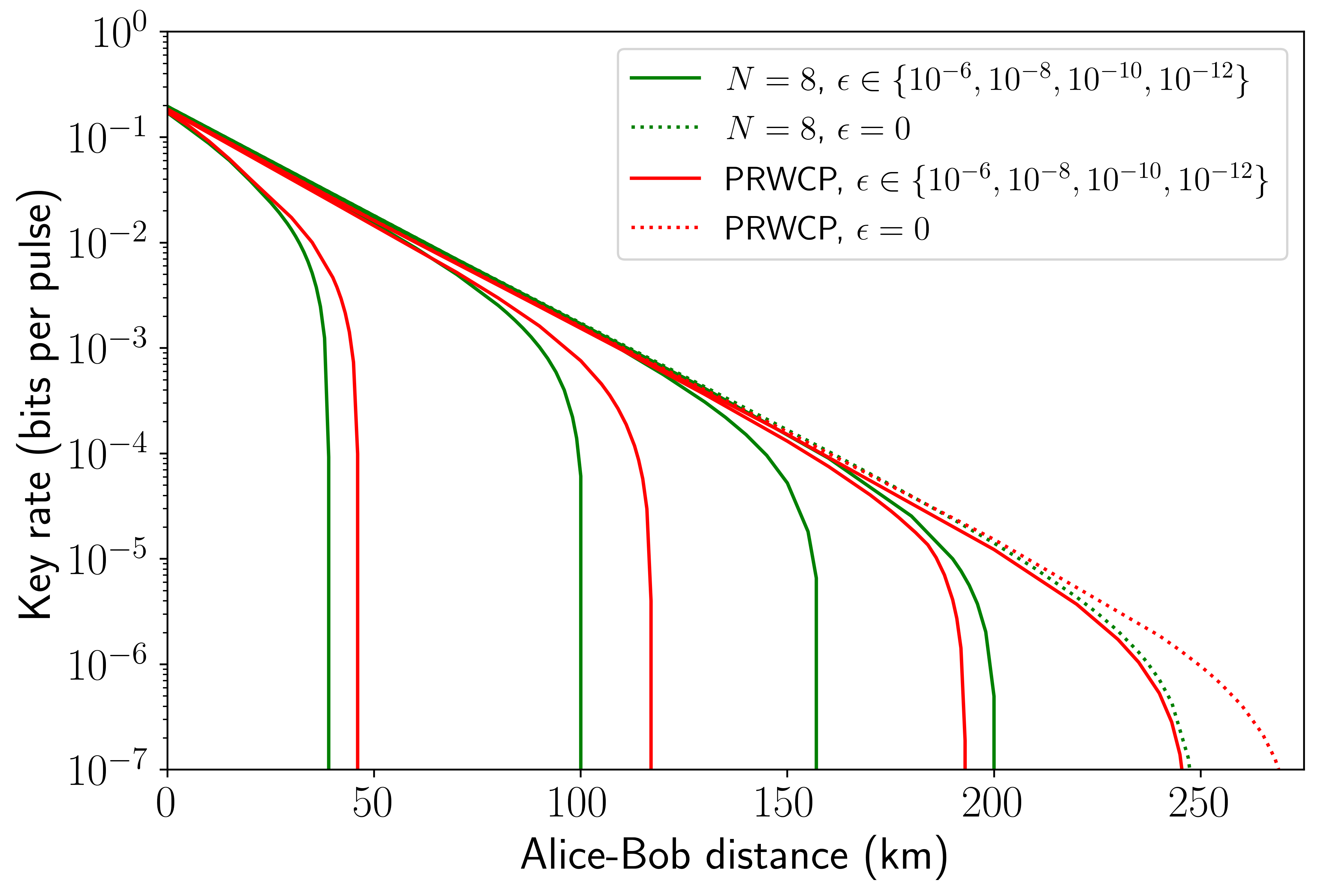}
\caption{\label{fig:leakage_results} Asymptotic secret-key rate vs distance for different values of $\epsilon$ for the two scenarios presented in \cref{sec:protocol}, $i.e$, active phase randomization (green lines) and passive phase randomization (red lines). In the case of a THA, $\epsilon$ corresponds to the magnitude of the back-reflected light $I$. The parameter $N=8$ refers to the number of discrete phases considered in the active phase randomization scenario. For both cases, we use the channel model presented in \cref{app:channel_model}.}
\end{figure}

The results are shown in \cref{fig:leakage_results}, where we plot the secret-key rate for various values of $\epsilon$ and the two practical scenarios introduced in~\cref{sec:protocol}. We find that it is possible to distill a secret key for values of $\epsilon$ up to about $\epsilon=10^{-6}$ in both scenarios. The considerable drop in performance when $\epsilon$ is large seems to be primarily due to the looseness of the bound on the photon number statistics given by \cref{eq:eigen_bounds} when $n$ and $\epsilon$ are large. This is because for large $n$ the magnitude of $p_{n|\beta}$ is similar to that of the perturbation parameter $\kappa$, especially for the weaker intensities (see \cref{app:use} for more details). Also, the bound on the inner products in \cref{cs_text} becomes less tight as $n$ and $\epsilon$ increase which is equivalent to consider more orthogonal states, which would allow Eve to perform, for instance, an unambiguous state discrimination attack \cite{USD, USD2}. Importantly, however, in some practical scenarios ---as $e.g.$, in a THA (see \cref{sec:practical})--- the value of $\epsilon$ can often be made as small as desired by simply increasing the isolation of Alice's lab.

\section{Characterized side channels}\label{sec:practical}

A fundamental merit of the method introduced in \cref{sec:security} is that it does not require to characterize the side-channel state, which simplifies its applicability in many practical scenarios. Precisely, as already discussed, Alice and Bob only need to know an upper bound on the parameter $\epsilon$. Notably, however, if the state of the side-channel system is partially (or fully) characterized, this additional information can be readily incorporated into the security analysis to enhance the resulting performance. To illustrate this point, in this section we consider such scenario. This might happen, for instance, when the information leakage is actively induced by Eve via a THA, and Alice and Bob are able to determine the state of Eve's back-reflected light. For example, if Alice deploys a single-mode optical fiber, then it is reasonable to assume that the state of the back-reflected light is in a single mode.

The security of QKD against THAs has been addressed in several different works \cite{tha3,tha2,tha1, tha4} and some of them assume that the state of Eve's back-reflected light is a coherent state (see \cite{tha3,tha2,tha4}). However, none of these studies consider simultaneous information leakage about all state preparation settings. In particular, information leakage about the global phase of the emitted pulses has been so far always disregarded. For instance, in the case of active phase randomization, Alice's phase modulator directly imprints the global phase onto Eve's injected pulses before they are reflected back to the channel. Here, we show how to adapt our security analysis to address the problem of information leakage with a characterized side channel. 

Importantly, before proceeding to the specific examples, it is worth remarking, as already mentioned, that in a THA the value of $\epsilon$ can be readily linked to the isolation of the transmitter. To see this, note that, as explained in \cref{sec:protocol}, the state that includes the side channel can always be written as shown in \cref{eq:dos}. 

Now, suppose that Alice places at the input of her transmitter a device that limits the maximum optical power of Eve’s injected light ---like \textit{e.g.}, an optical fuse \cite{fuse}, or a power limiter \cite{limiter}--- to, say, 1 dBm. In addition, suppose that the total attenuation suffered by Eve's Trojan pulses in their whole round trip (in and out of Alice's transmitter) is say 140dB, and that these pulses have a temporal width of 1 ns and a wavelength of 1550 nm. Then, the average number of photons contained in the side-channel system $E$ is limited to approximately $10^{-8}$ photons per round. This corresponds to $\epsilon=10^{-8}$ in \cref{fig:leakage_results} since $\epsilon\leq I$ \cite{pereira2023}, where $I$ denotes the maximum intensity of Eve's back-reflected light in a THA.

\subsection{THA against active phase randomization}\label{sec:tha1}

\begin{figure}[htbp]
    \centering
    \includegraphics[width= 8.6cm, height=2cm]{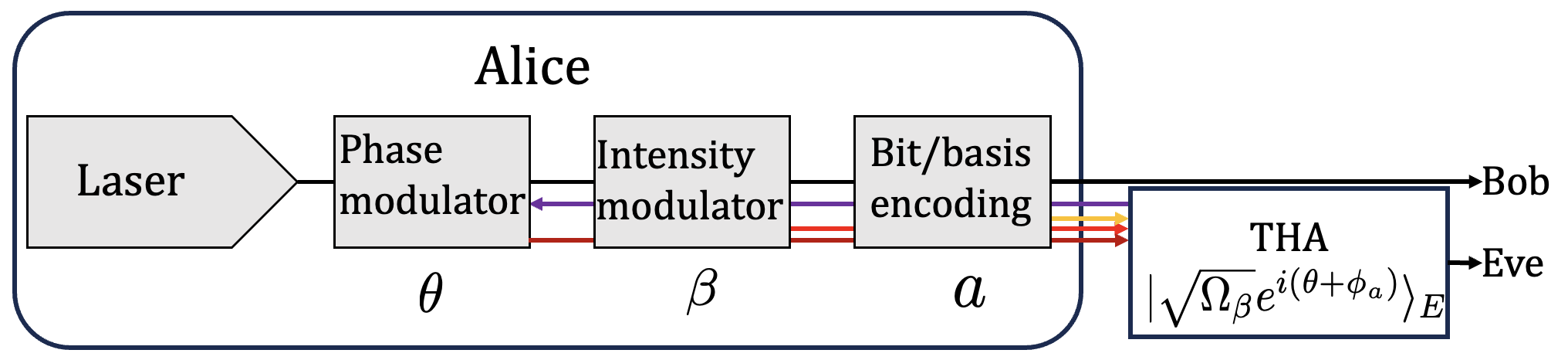}
    \caption{Graphical representation of a THA in which Eve injects strong coherent light into Alice's transmitter and the output system $E$ is a coherent state that depends on the selected settings $\beta$, $\theta$ and $a$. Precisely, the intensity is modulated as that of the legitimate signals and its phase is the sum of the global phase and an extra phase that depends on the bit and basis selection. Importantly, here we consider that the state of the side channel is fully characterized.}\label{fig:THA_1}
\end{figure}

First, we consider the scenario illustrated in \cref{fig:THA_1} in which Alice's transmitter applies perfect discrete phase randomization. That is, $g(\theta)$ is characterized by \cref{eq:pdf}. The case of passive phase randomization could be studied similarly by simply replacing the PDF $g(\theta)$ and applying the techniques in \cite{phase2}. Moreover, similarly to \cite{tha3,tha2,tha1, tha4}, we shall assume a simple model for Eve's THA. In particular, we shall consider that she injects coherent light into Alice's transmitter and the state of her back-reflected pulses, represented by $\ket*{\lambda_{\beta,a,\theta}}_E$ in \cref{eq:THA_generalstate}, is fully characterized and given by the coherent state $\ket*{\sqrt{\Omega_{\beta}}e^{i(\theta+\phi_a)}}_{E}$. That is, Eve's coherent pulses, after being reflected, are modulated by Alice's optical modulators just like Alice's signals. Notably, this implies that $\ket*{\lambda_{\beta,a,\theta}}_E$ can be decomposed as 
\begin{equation}
\ket*{\lambda_{\beta,a,\theta}}_E = e^{-\frac{\Omega_{\beta}}{2}}\ket*{0}_{E}+e^{-\frac{\Omega_{\beta}}{2}}\sum_{n=1}^{\infty}  \frac{(\sqrt{\Omega_{\beta}}e^{i(\theta+\phi_a)})^{n}}{\sqrt{n!}}\ket*{n}_{E}.
\end{equation}
This means that $\epsilon$ can be readily linked to the intensity of the output state as 
\begin{equation}
\label{eq:link}
\epsilon=\max_{\beta}\Big\{1-e^{-\Omega_{\beta}}\Big\}.
\end{equation}

Moreover, we have that the intensity $\Omega_{\beta}$ of Eve's pulses is modulated by Alice's IM that encodes the different intensities of the decoy-state protocol, thus satisfies \cite{tha2}
\begin{equation}
\label{eq:int_modu}
\Omega_{\beta}=\frac{\beta}{\mu}I,
\end{equation}
where $I$ represents the maximum intensity of the back-reflected light and $\mu$ the signal intensity. This way it is guaranteed that 
$\frac{\Omega_{\gamma}}{\Omega_{\zeta}}=\frac{\gamma}{\zeta}$,
with $\gamma,\zeta\in\{\mu,\nu,\omega\}$. Intuitively this amounts to say that the intensity of Eve's pulses is attenuated exactly as the legitimate signals.

In addition, the global phase $\theta+\phi_a$ of Eve's back-reflected light is equal to the sum of the global phase $\theta$ of Alice's signals and a phase $\phi_a$ that depends on the different BB84 encoded states. In particular, we shall consider that 
\begin{equation}\label{eq:phases_states}
\phi_{0_Z}=0,\quad \phi_{1_Z}=\pi,\quad \phi_{0_X}=\pi/2\quad \text{and}\quad \phi_{1_X}=3\pi/2. 
\end{equation}

To apply the formalism introduced in \cref{sec:security} to this scenario, one needs to estimate a lower bound on the single-photon yield $Y^{1}_{\mu, Z}$ and an upper bound on the bit-error rate $e^{{\rm b}, 1}_{\beta, X}$ via~\cref{eq:min,eq:min_phase,eq:phase_error_bound} together with the linear programs introduced in~\cref{LP_final,LP_final_error}. These linear programs require to estimate the fidelity $F(\ket*{n^{\epsilon}_{\gamma,a, h(\theta)}},\ket*{{n^{\epsilon}_{\zeta, a, h(\theta)}}_{BE}})$, where here the states $\ket*{n^{\epsilon}_{\beta,a, h(\theta)}}_{BE}$ play the role of the states $\ket*{n^{\epsilon}_{\beta, a, g(\theta)}}_{BE}$ introduced in~\cref{eq:states}. Importantly, the unnormalized signals are now perfectly characterized and given by \cite{phase0}
\begin{equation}
\begin{aligned}\label{eq:nstates_Ex1}
\ket*{\bar{n}^{\epsilon}_{\beta, a, h(\theta)}}_{BE} =& \sum_{l=0}^{N-1} e^{-i\frac{2\pi}{N}nl}\hat{V}^{a}_{B}\ket*{\sqrt{\beta}e^{i\frac{2\pi}{N}l}}_B\\
&\otimes\ket*{\sqrt{\Omega_{\beta}}e^{i(\frac{2\pi}{N}l+\phi_a)}}_E.
\end{aligned}
\end{equation}
where the bar in $\ket*{\bar{n}^{\epsilon}_{\beta, a, h(\theta)}}_{BE}$ denotes that the state is not normalized.

By combining \cref{eq:nstates_Ex1} with the fact that the inner product between two coherent states satisfies $\braket{\beta}{\alpha}=e^{-\frac{1}{2}\left(|\beta|^2+|\alpha|^2-2 \beta^* \alpha\right)}$, one can straightforwardly compute the required fidelity and it is not necessary to use the SDP techniques presented in \cref{sec:inner_SDP}. As for the probabilities $p^{\epsilon}_{n|\beta,a}$, we have that \cite{phase0}
\begin{equation} 
p^{\epsilon}_{n|\beta, a} =\frac{\braket*{\bar{n}^{\epsilon}_{\beta, a, h(\theta)}}{\bar{n}^{\epsilon}_{\beta, a, h(\theta)}}_{BE}}{\sum_{m=0}^{N-1}\braket*{\bar{m}^{\epsilon}_{\beta, a, h(\theta)}}{\bar{m}^{\epsilon}_{\beta, a, h(\theta)}}_{BE}}.
\end{equation}
Since these statistics are known precisely now, the gain inequalities in~\cref{LP_final,LP_final_error} can be replaced by equalities if $n_{\text{cut}}=N-1$, and the bound in \cref{eq:bound_main} is not needed.

Similarly, the fidelity in~\cref{eq:basis_dependence}, used to obtain the quantum-coin imbalance can now be calculated as 
\begin{equation}\label{eq:basis_dependence2}
\begin{split}
&F_{\beta}^{n,\epsilon} = |\braket*{\Psi^{n,\epsilon}_{\beta, Z, h(\theta)}}{\Psi^{n,\epsilon}_{\beta, X, h(\theta)}}_{ABE}|^2,
\end{split}
\end{equation}
%
where the states $\ket*{\Psi^{n, \epsilon}_{\beta, \alpha, h(\theta)}}_{ABE}$ are defined in~\cref{eq:bas_dep_states}. For simplicity, we compute \cref{eq:basis_dependence2} numerically. Note that, since all  the necessary inner products to compute \cref{eq:basis_dependence2} are now known, here we do not need to apply the triangle inequality approach of the previous section to estimate $F_{\beta}^{n, \epsilon}$. Finally, given the upper bounds on $e^{{\rm b}, 1}_{\beta, X}$ and the quantum-coin imbalance parameter, the phase-error rate $e^{{\rm ph}, 1}_{\mu}$ can be upper bounded with~\cref{qwe}, and the secret-key rate can be computed using~\cref{eq:skr_general}. 

\begin{figure}[htbp]
\centering\includegraphics [width= 8.6cm, height=6cm] {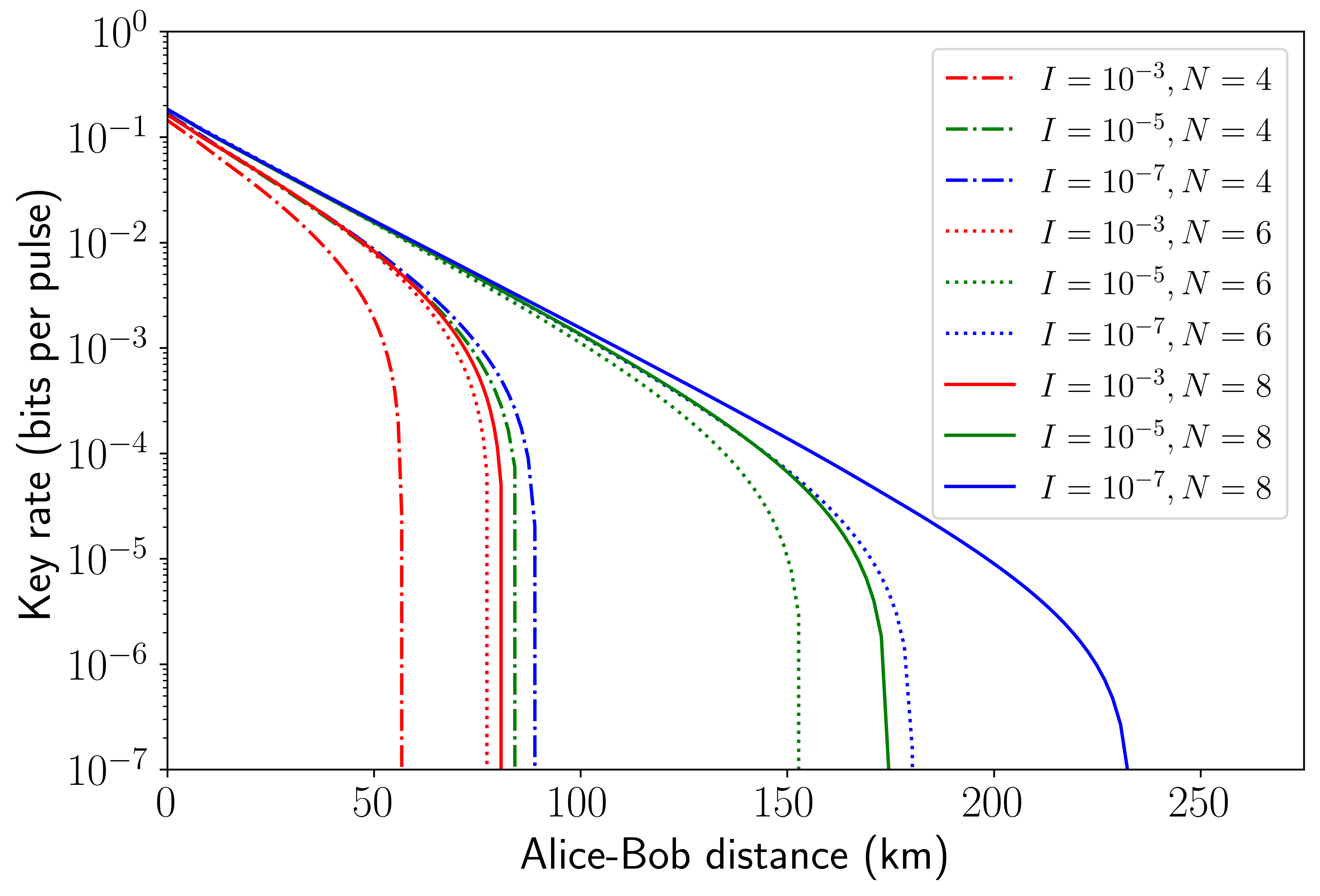}
\caption{\label{fig:THA_1_results}Secret-key rate vs distance for different values of the maximum intensity $I$ of Eve's back-reflected light in a THA. The parameter $N$ refers to the total number of random global phases imprinted by the phase modulator. Once again, we use the channel model presented in \cref{app:channel_model} and optimize the values of the intensities $\mu$ and $\nu$ to maximize the secret-key rate. In contrast to \cref{fig:leakage_results}, here the state of the side channel is characterized.}
\end{figure}

\begin{figure}[htbp]
\centering\includegraphics [width= 8.6cm, height=6cm] {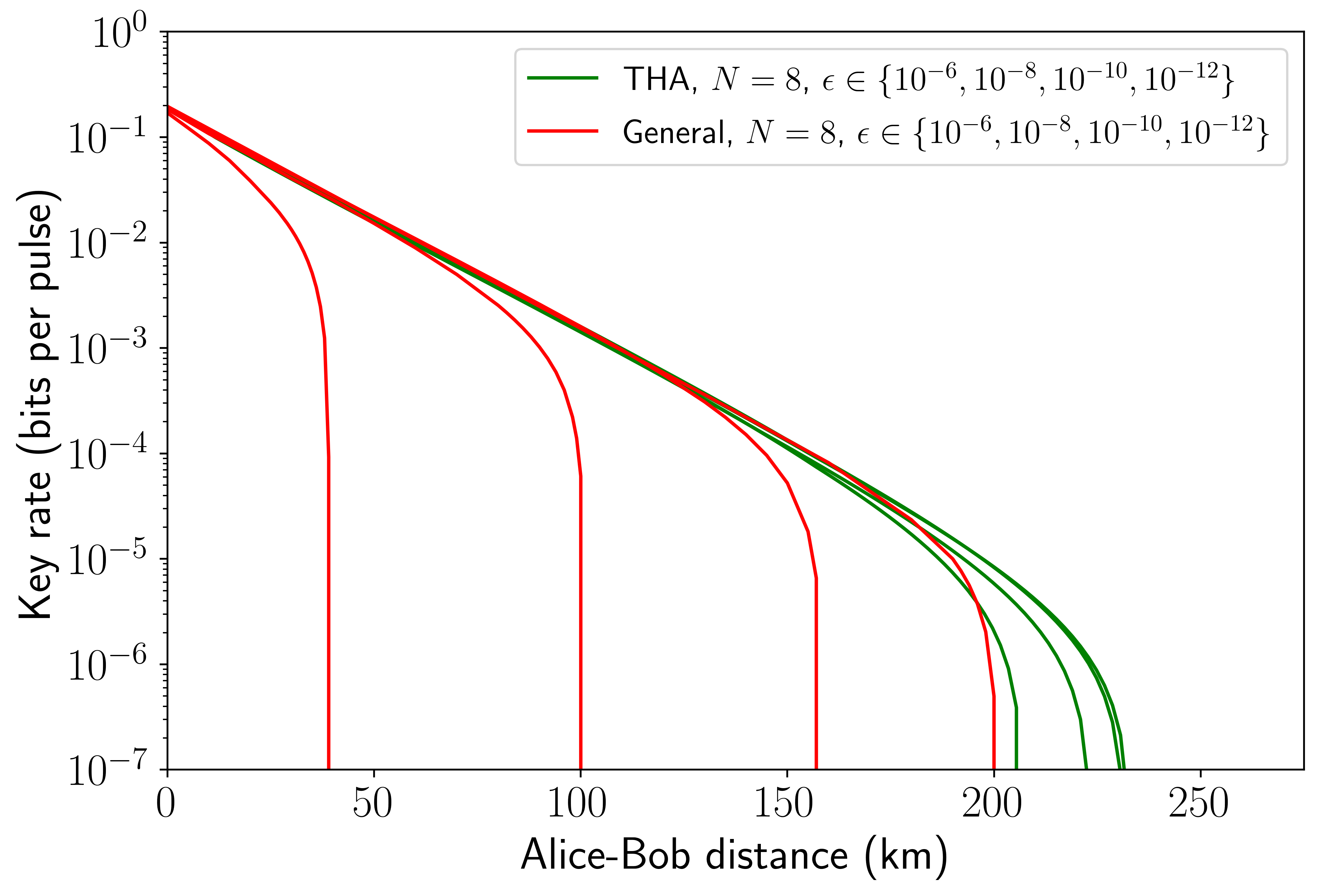}
\caption{\label{fig:THA_1_extra} Comparison of the secret key rate for the scenarios presented in \cref{sec:security} named as "General" and \cref{sec:tha1} (THA) for different values of $\epsilon$. This parameter is related to the maximum intensity $I$ of Eve's back-reflected light in a THA, as expressed in \cref{eq:link}. The parameter $N$ refers to the total number of random global phases imprinted by the phase modulator.}
\end{figure}

The results are shown in \cref{fig:THA_1_results}, where we plot the secret-key rate for different values of the maximum intensity $I$ of Eve's back-reflected light, as well as different numbers $N$ of the random phases used for global phase randomization. In the simulations, we optimize the secret-key rate over the intensities $\mu$ and $\nu$ and set the vacuum intensity again to $\omega=0$. It is clear from the figure that employing just $N=4$ global phase values results in poor performance, regardless of the actual intensity of Eve's back-reflected light, which is consistent with the results presented in \cite{phase0} in the absence of information leakage.

Finally, \cref{fig:THA_1_extra} presents a comparison between the scenarios in \cref{sec:security,sec:tha1} using \cref{eq:link} to link the parameter $\epsilon$ with the maximum intensity of Eve's back-reflected light $I$. As expected, we find that incorporating additional information about the side-channel state drastically improves the secret key rate.

\subsection{Active phase randomization as a countermeasure against THAs}

Besides optical isolation, another potential countermeasure against THAs is to add an extra phase modulator (PM) to Alice's transmitter to blur the information contained in the phase of Eve's back-reflected light~\cite{gissin,tha2,tha3}. However, a rigorous quantitative analysis of its effectiveness has remained elusive so far, particularly because Eve's THA will also provide her with partial information about the value of such extra phase. Here, we investigate this scenario by considering a second side-channel system in a coherent state $\ket*{\sqrt{\lambda} e^{i\phi}}_{E_2}$ with fixed intensity $\lambda\leq I_l$, and whose phase is determined by the phase $\phi$ of the extra PM. For simplicity, in this subsection we shall consider that the THA targets only the bit/basis encoder and the intensity modulator. That is,  unlike the previous subsection, here we assume that Eve obtains no information about the global phase. This is because in a THA against a transmitter that uses passive phase randomization, the intensity of Eve's injected light entering Alice's laser cavity is typically too weak to influence the global phase of the emitted pulses due to its optical isolation. Still, we remark that the scenario with information leakage about the global phases could be addressed with the same tools presented in previous sections.

\begin{figure}[htbp]
    \centering
    \includegraphics[width= 8.6cm, height=2cm]{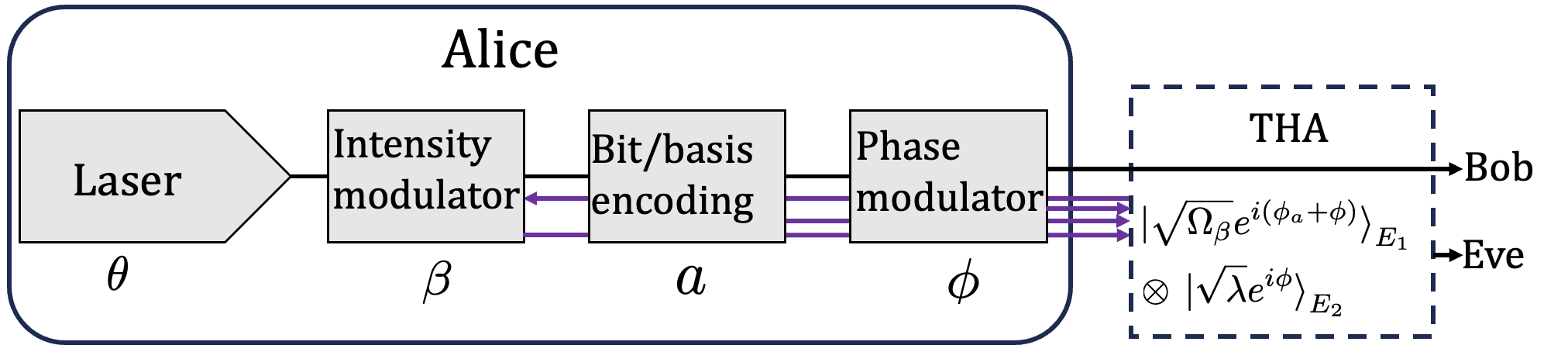}
    \caption{Analysis of the effectiveness of using an additional PM as a countermeasure against THAs. For concreteness, we consider that the phase randomization process is carried out by operating the laser in gain switching. Eve's back-reflected light is described by two coherent states $\ket*{\sqrt{\Omega_{\beta}}e^{i(\phi+\phi_{a})}}_{E_1}\otimes\ket*{\sqrt{\lambda}e^{i\phi}}_{E_2}$. When Eve performs a THA, some of the light is reflected by the laser and goes through the intensity modulator, the bit/basis encoder and the extra PM, which corresponds to system $E_1$, while other photons are reflected by the bit/basis encoder and go only through the additional PM, which corresponds to system $E_2$.}\label{fig:THA_2}
\end{figure}

The scenario is illustrated in \cref{fig:THA_2}. We assume that the global phase $\theta$ of the transmitted states is imprinted by gain-switching the laser. In addition, for simplicity, we consider that such phase is perfectly random, that is, it follows an uniform distribution $f(\theta)=(2\pi)^{-1}$, although our analysis can be applied to any given $g(\theta)$. As for the side channel, we assume that Eve's back-reflected light is in a coherent state $\ket*{\sqrt{\Omega_{\beta}}e^{i(\phi+\phi_{a})}}_{E_1}\otimes\ket*{\sqrt{\lambda}e^{i\phi}}_{E_2}$  where, similarly to~\cref{eq:int_modu}, $\Omega_{\beta}$ is given by $\Omega_{\beta}=\beta I/\mu$. 

Precisely, if there is no information leakage, the state prepared by Alice in a particular round is simply given by 
\begin{equation}
\label{eq:ideal2}
\begin{aligned}
&\rho_{f(\theta),g_{e}(\phi),BE}^{\beta,a,{\rm ideal}}\\
&=\int_{0}^{2\pi}\!\!\int_{0}^{2\pi}\frac{g_e(\phi)}{2\pi}\hat{P}\left[\hat{V}_B^a
\ket*{\sqrt{\beta}e^{i(\theta+\phi)}}_B\ket{\rm v}_{E}\right] d\theta d\phi\\
&=\sum_{n=0}^{\infty}{p}_{n|\beta}\PP{\hat{V}_B^a\ket{n}_B}\otimes\dyad{\rm v}_{E},
\end{aligned}
\end{equation}
where $E\equiv E_1E_2$, the probability $p_{n|\beta}=e^{-\beta}\beta^{n}/n!$, the state $\ket{n}_B$ represents an $n$-photon Fock state, and $g_e(\phi)$ is the PDF of the extra PM. On the other hand, if we account for information leakage, we have that
\begin{equation}
\label{eq:THA2}
\begin{aligned}
\rho_{f(\theta), g_e(\phi),BE}^{\epsilon,\beta,a}=&
\int_{0}^{2\pi}\!\!\int_{0}^{2\pi}\frac{g_e(\phi)}{2\pi}\hat{P}\Big[\hat{V}_B^a
\ket*{\sqrt{\beta}e^{i(\theta+\phi)}}_B\\
&\otimes\ket*{\sqrt{\Omega_{\beta}}e^{i(\phi+\phi_{a})}}_{E_{1}}
\ket*{\sqrt{\lambda}e^{i\phi}}_{E_{2}}\Big] d\theta d\phi.
\end{aligned}
\end{equation}

Importantly, since by assumption no information about $\theta$ is leaked to Eve in this scenario, this state can be rewritten as a tensor product of Bob's and Eve's systems, \textit{i.e.}, $\rho_{f(\theta), g_e(\phi),BE}^{\epsilon,\beta,a}=\sum_{n=0}^{\infty}p_{n|\beta}\dyad{n_{a}}{n_{a}}_{B}\otimes\rho_{g_{e}(\phi) ,E}^{\beta,a,{\rm leak}}$. Here, the states $\ket{n_a}_B=\hat{V}_B^{a}\ket{n}_B$ can be written as $\ket{n_{0_Z}}_B=|n\rangle_{B_1}|0\rangle_{B_2}, \ket{n_{1_Z}}_B=|0\rangle_{B_1}|n\rangle_{B_2}$, and
\begin{equation}\label{eq:purificatinos_basis_dep}
\begin{aligned}
&\ket{n_{0_X}}_B=\sum_k \frac{1}{\sqrt{2^n}} \sqrt{\left(\begin{array}{c}
n \\
k
\end{array}\right)}|n-k\rangle_{B_1}|k\rangle_{B_2}, \\
&\ket{n_{1_X}}_B=\sum_k(-1)^k \frac{1}{\sqrt{2^n}} \sqrt{\left(\begin{array}{l}
n \\
k
\end{array}\right)}|n-k\rangle_{B_1}|k\rangle_{B_2},
\end{aligned}
\end{equation}
where $B_1$ and $B_2$ represent two optical modes associated with the particular type of encoding. For example, in polarization encoding, $B_1$ ($B_2$) may represent the horizontal (vertical) polarization mode, whereas in time-bin encoding it may represent the early (late) time bin. On the other hand, the side-channel state $\rho_{g_{e}(\phi) ,E}^{\beta,a,{\rm leak}}$ is defined as 
\begin{equation}
\begin{aligned}\label{eq:matrix_side}
&\rho_{g_{e}(\phi) ,E}^{\beta,a,{\rm leak}}\\
&=\int_{0}^{2\pi}g_e(\phi)\hat{P}\Big[\ket*{\sqrt{\Omega_{\beta}}e^{i(\phi+\phi_{a})}}_{E_{1}}
\ket*{\sqrt{\lambda}e^{i\phi}}_{E_{2}}\Big] d\phi.
\end{aligned}
\end{equation}

To apply our security analysis to this scenario, we need to lower bound the fidelity between purifications of the states $\rho_{n,f(\theta), g_e(\phi),BE}^{\epsilon,\beta,a}=\dyad{n_{a}}{n_{a}}_{B}\otimes\rho_{g_{e}(\phi) ,E}^{\beta,a,{\rm leak}}$ with different intensity settings. Due to the tensor product structure of these states, it holds that $F(\ket*{\Psi^{n,\epsilon}_{\zeta,a, f(\theta),g_{e}(\phi)}}_{PBE},\ket*{\Psi^{n, \epsilon}_{\gamma,a, f(\theta),g_{e}(\phi)}}_{PBE})=F(\ket*{\Psi^{\text{leak}}_{\zeta,a, g_e(\phi)}}_{PE}, \ket*{\Psi^{{\rm leak}}_{\gamma,a, g_e(\phi)}}_{PE})$, where the state $\ket*{\Psi^{n,\epsilon}_{\beta,a, f(\theta),g_{e}(\phi)}}_{PBE}$ represents a purification of $\rho_{n,f(\theta), g_e(\phi)BE}^{\epsilon,\beta,a}$ and $\ket*{\Psi^{\text{leak}}_{\beta,a, g_e(\phi)}}_{PE}$ represent a purification of $\rho_{g_{e}(\phi) ,E}^{\beta,a,{\rm leak}}$, being $P$ the purification system. To compute this latter quantity we assume for simplicity that $g_{e}(\phi)$ is given by \cref{eq:pdf}, $i.e.$, the external PM imprints one of  $N$ evenly distributed phases in each pulse with equal probability. Because the side-channel states in \cref{eq:matrix_side} are completely characterized, we can directly compute the fidelity between two  density matrices $\rho$ and $\sigma$ as $F(\rho, \sigma)=(\operatorname{Tr} \sqrt{\sqrt{\rho} \sigma \sqrt{\rho}})^2$. Note that
\begin{equation}\label{eq:keyalert}
\begin{aligned}
&F(\rho_{g_{e}(\phi) ,E}^{\zeta,a,{\rm leak}},\rho_{g_{e}(\phi) ,E}^{\gamma,a,{\rm leak}})\\
&=\max_{\substack{\ket*{\Psi^{\text{leak}}_{\zeta,a,g_e(\phi)}}_{PE} \\ \ket*{\Psi^{\text{leak}}_{\gamma,a,g_e(\phi)}}_{PE} }} F(\ket*{\Psi^{\text{leak}}_{\zeta,a,g_e(\phi)}}_{PE}, \ket*{\Psi^{{\rm leak}}_{\gamma,a,g_e(\phi)}}_{PE}),    
\end{aligned}
\end{equation}
where the maximization runs over all possible purifications. In general, however, \cref{eq:keyalert} cannot be computed numerically, as the states are infinitely dimensional. To solve this, we use the triangle inequality of the Bures distance, in similar fashion to \cref{eq:triangle}. Now,
\begin{equation}
\begin{aligned}
&d_{B}(\rho_{g_{e}(\phi) ,E}^{\zeta,a,{\rm leak}},\rho_{g_{e}(\phi) ,E}^{\gamma,a,{\rm leak}})\leq d_{B}(\rho_{g_{e}(\phi) ,E}^{\zeta,a,{\rm leak}},\rho_{g_{e}(\phi) ,E}^{M,\zeta,a,{\rm leak}})+\\
&d_{B}(\rho_{g_{e}(\phi) ,E}^{M,\zeta,a,{\rm leak}},\rho_{g_{e}(\phi) ,E}^{M,\gamma,a,{\rm leak}})+d_{B}(\rho_{g_{e}(\phi) ,E}^{M,\gamma,a,{\rm leak}},\rho_{g_{e}(\phi) ,E}^{\gamma,a,{\rm leak}}),
\end{aligned}  
\end{equation}
where $\rho_{g_{e}(\phi) ,E}^{M,\beta,a,{\rm leak}}$ represents a finite projection  of $\rho_{g_{e}(\phi) ,E}^{\beta,a,{\rm leak}}$ onto the space with up to $M$ photons. The quantity $d_{B}(\rho_{g_{e}(\phi) ,E}^{M,\zeta,a,{\rm leak}},\rho_{g_{e}(\phi) ,E}^{M,\gamma,a,{\rm leak}})$ can be obtained numerically, while $d_{B}(\rho_{g_{e}(\phi) ,E}^{\beta,a,{\rm leak}},\rho_{g_{e}(\phi) ,E}^{M,\beta,a,{\rm leak}})$ is calculated in \cref{app:use}.

After obtaining a lower bound on $F(\rho_{g_{e}(\phi) ,E}^{\zeta,a,{\rm leak}},\rho_{g_{e}(\phi) ,E}^{\gamma,a,{\rm leak}})$ one can run the linear program presented in \cref{LP_final} to estimate the single-photon yield. 

To obtain an upper bound on the phase-error rate, on the other hand, one needs to determine the quantum coin imbalance of the detected coin rounds defined in \cref{eq:basis_dependence}, which depends on the fidelity $F^{n, \epsilon}_{\beta}$ between the states $|\Psi^{n,\epsilon}_{\beta, \alpha, f(\theta), g_e(\phi)}\rangle_{PABE}$ with different bases, defined similarly to those in \cref{eq:bas_dep_states}, $i.e.$,
\begin{equation}
\begin{aligned}
\ket*{\Psi^{n,\epsilon}_{\beta, \alpha, f(\theta), g_{e}(\phi)}}_{PABE}=&\frac{1}{\sqrt{2}}
\sum_{j=0}^{1} \ket{j_\alpha}_A\ket*{n^{\epsilon}_{\beta,j_\alpha, f(\theta), g_{e}(\phi)}}_{PBE}.
\end{aligned}
\end{equation}

To calculate $F_{\beta}^{n,\epsilon}$, we note that, for this particular scenario, the states $\ket*{n^{\epsilon}_{\beta,a, f(\theta), g_e(\phi)}}_{PBE}$ take the simple form
\begin{equation}
\ket*{n^{\epsilon}_{\beta,a, f(\theta), g_e(\phi)}}_{PBE}=\ket{n_{a}}_{B}\otimes\ket*{\Psi_{\beta,a, g_e(\phi)}^{\text{leak}}}_{PE}.
\end{equation}
Thus we have that  
\begin{equation}\label{eq:basis_dependence22}
\begin{aligned}
F_{\beta}^{n,\epsilon} &= \big|\braket*{\Psi^{n,\epsilon}_{\beta,X,f(\theta),g_e(\phi)}}{\Psi^{n,\epsilon}_{\beta,Z,f(\theta),g_e(\phi)}}_{PABE}\big|^2\\
&=\bigg|\frac{1}{2\sqrt{2^{n+1}}}\big(\braket*{\Psi^{\text{leak}}_{\beta, 0_X,g_e(\phi)}}{\Psi^{\text{leak}}_{\beta, 0_Z,g_e(\phi)}}_{PE}\\
&+\braket*{\Psi^{\text{leak}}_{\beta, 1_X,g_e(\phi)}}{\Psi^{\text{leak}}_{\beta, 0_Z,g_e(\phi)}}_{PE}\\
&+\braket*{\Psi^{\text{leak}}_{\beta, 0_X,g_e(\phi)}}{\Psi^{\text{leak}}_{\beta, 1_Z,g_e(\phi)}}_{PE}\\
&-(-1)^n\braket*{\Psi^{\text{leak}}_{\beta, 1_X,g_e(\phi)}}{\Psi^{\text{leak}}_{\beta, 1_Z,g_e(\phi)}}_{PE}\big)\bigg|^{2}.
\end{aligned}
\end{equation}

To compute the different terms in this equation, we cannot use the approach of \cref{eq:keyalert}, because it does not guarantee that the same purification is used for every ${\Psi^{\text{leak}}_{\beta, j_\alpha,g_e(\phi)}}_{PE} $. Therefore, here, we select the following purifications for the leakage states
\begin{equation}
\label{eq:puri_sc_2}
\begin{aligned}
\ket*{\Psi^{\text{leak}}_{\beta, a,g_e(\phi)}}_{PE}=& \frac{1}{\sqrt{N}}\sum_{j=0}^{N-1}|j\rangle_{P}|\sqrt{\Omega_{\beta}}e^{i(\phi_{j}+\phi_{a}+\delta_{a})}\rangle_{E_1}\\
&\otimes|\sqrt{\lambda}e^{i(\phi_{j}+\delta_{a})}\rangle_{E_2},
\end{aligned}
\end{equation}
where we can freely choose any $\delta_{a}$ for every bit/basis value $a$ in the set of $N$ possible phases, given by $\phi_j=\frac{2\pi j}{N}$. After this, we can optimize the value of $F^{n, \epsilon}_{\beta}$ numerically by finding the set of purifications ---or equivalently a set of $\delta_{a}$--- that maximizes \cref{eq:basis_dependence22}.


\begin{figure}[H]
\centering\includegraphics [width= 8.6cm, height=6cm] {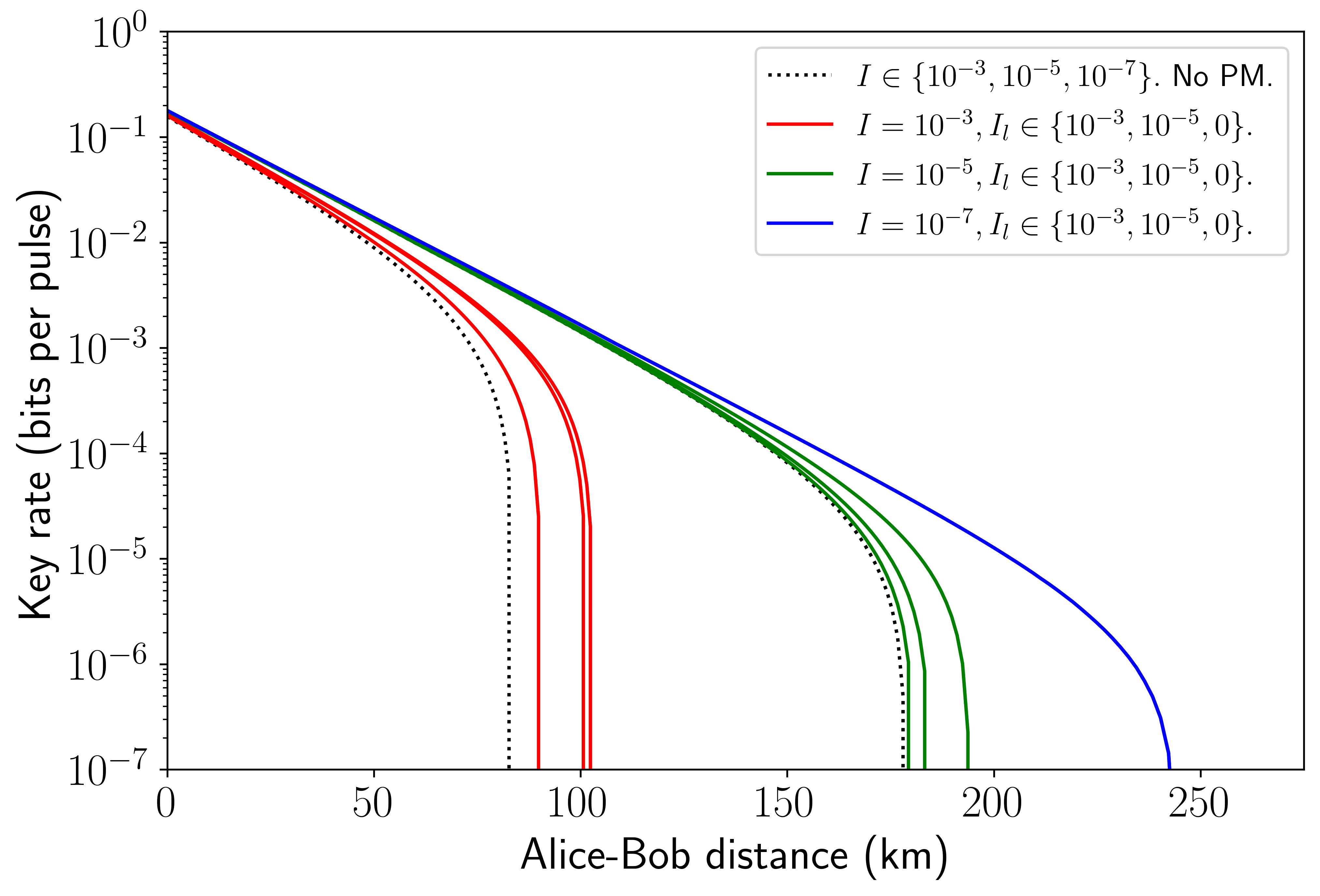}
\caption{\label{fig:THA_2_results} Secret-key rate vs distance for different values of the intensities $I$ and $I_l$. The dotted lines correspond to the scenario without an additional PM. Note that in general such PM helps to improve the key rate, and that this improvement becomes more noticeable if the PM is well isolated from Eve's THA, or equivalently, if $I_l$ is low. The case $I_l=0$ corresponds to perfect isolation. Note that, if $I=10^{-7}$ the four lines essentially overlap. Finally, for concreteness, here we assume $N=4$ random phases, and we have tested numerically that further increasing $N$ does not significantly improve the performance, at least in the parameter regime considered.}
\end{figure}

Having computed the basis dependence of the source, one combines \cref{LP_final_error} with \cref{eq:min_phase,eq:phase_error_bound} to bound the $X$-basis bit-error rate, and then applies \cref{qwe} to obtain the phase-error rate.

In \cref{fig:THA_2_results}, we plot the attainable secret-key rate for different values of the maximum intensities $I$ and $I_l$. As in previous sections, we optimize $\mu$ and $\nu$, and set $\omega=0$. This figure shows that the inclusion of an additional active randomization step indeed allows to mitigate the impact of a THA in comparison to the scenario in which that device is not present, even if it can only imprint $N=4$ different phases. We have also found that a further increase in $N$ does not alter the performance significantly for the values of $I$ and $I_l$ considered in the figure. The improvement in key rate is specially relevant if the PM can be well isolated from Eve's THA, $i.e.$, if $I_l$ is low.
Moreover, as expected, when the information leaked from the encoding and intensity modulators is very low, regardless of the amount of information that Eve could obtain about the extra phase $\phi$, the secret-key rate is not significantly altered.

\section{Conclusions}\label{sec:conclusions}

In this paper, we have addressed the problem of information leakage in quantum key distribution (QKD). Importantly, unlike previous analyses, our security proof considers leakage from every setting selected at the source. Besides, it holds for any phase distribution function and, consequently, it does not require the stringent condition that the source produces perfect phase-randomized weak coherent pulses. In addition, using the proof in a real-world experiment only demands the measurement of one parameter, which is linked to the isolation of the source  and so it simplifies its applicability. Furthermore, the secret key rate simulations suggest that the fact that Eve could learn partial information about the global phase of the laser pulses is very detrimental in terms of performance, a problem that has been overlooked in previous studies.

While the security proof introduced requires minimal characterization of the state of the side channel, it can also be adopted to incorporate more information about such state. Indeed, our findings suggest that the performance of a QKD protocol under realistic conditions could be severely improved with precise side-channel characterization. Finally, as a side point, we have quantified the effectiveness of including an additional external phase modulator for phase-randomization as a countermeasure against Trojan horse attacks. This countermeasure has been proposed in various previous studies but its quantitative evaluation has been elusive so far.


\section{Acknowledgements}\label{Acknowledgements}
We acknowledge insightful discussions with Marco Lucamarini. This work was supported by the Galician Regional Government (consolidation of Research Units: AtlantTIC), the Spanish Ministry of Economy and Competitiveness (MINECO), the Fondo Europeo de Desarrollo Regional (FEDER) through the grant No. PID2020-118178RB-C21, MICIN with funding from the European Union NextGenerationEU (PRTR-C17.I1) and the Galician Regional Government with own funding through the “Planes Complementarios de I+D+I con las Comunidades Autónomas” in Quantum Communication, the European Union’s Horizon Europe Framework Programme under the Marie Sklodowska-Curie Grant No. 101072637 (Project QSI) and the project ”Quantum Security Networks Partnership” (QSNP, grant agreement No 101114043). X.S. acknowledges support from an FPI predoctoral scholarship granted by the Spanish Ministry of Science, Innovation and Universities. M.C. acknowledges support from a ``Salvador de Madariaga'' grant from the Spanish Ministry of Science, Innovation and Universities, grant No. PRX22/00192. K.T. acknowledges support from
JSPS KAKENHI Grant Number 23H01096.

\section{Data availability statement}

No new data were created or analyzed in this study.

\section{Author contributions}

M.C. identified the need for the research project. X.S. and A.N. developed the security proof with help from all authors and X.S performed the numerical simulations. X.S., A.N. and M.C. wrote the manuscript, and all authors contributed towards improving it and checking the validity of the results.

\appendix

\section{Quantum coin equivalences}

As stated in the main text, computing the secret key rate requires to find suitable bounds on the parameters $Y_{\mu, Z}^{1}$ and $e^{{{\rm ph}, 1}}_{\mu}$. In this Appendix, we elaborate on the equivalence between the actual protocol and a fictitious scenario in which quantum coins are generated, which allows to find such bounds.

\subsection{Yield estimation}\label{app:CS}
Here we justify the use of the inequality given  in~\cref{cs_text}.
For this, we consider the quantum-coin argument introduced in~\cite{GLLP, lo_preskill} to demonstrate the security of QKD with imperfect devices. This method has been recently refined in \cite{Guille_framework} to prove the finite-key security of QKD against coherent attacks in the presence of arbitrary encoding flaws and side channels in protocols that consider single photon sources. This latter analysis relies on a random tagging assignment that allows to establish an equivalence between the actual protocol and a virtual protocol in which Alice prepares quantum-coin states. In our scenario, however, the situation is simpler than that considered in \cite{Guille_framework}, since here we focus on the asymptotic-key regime, and thus defining an efficient random tagging assignment is not crucial.

\begin{figure}[H]
\centering\includegraphics [width= 8.8cm, height=10.1cm] {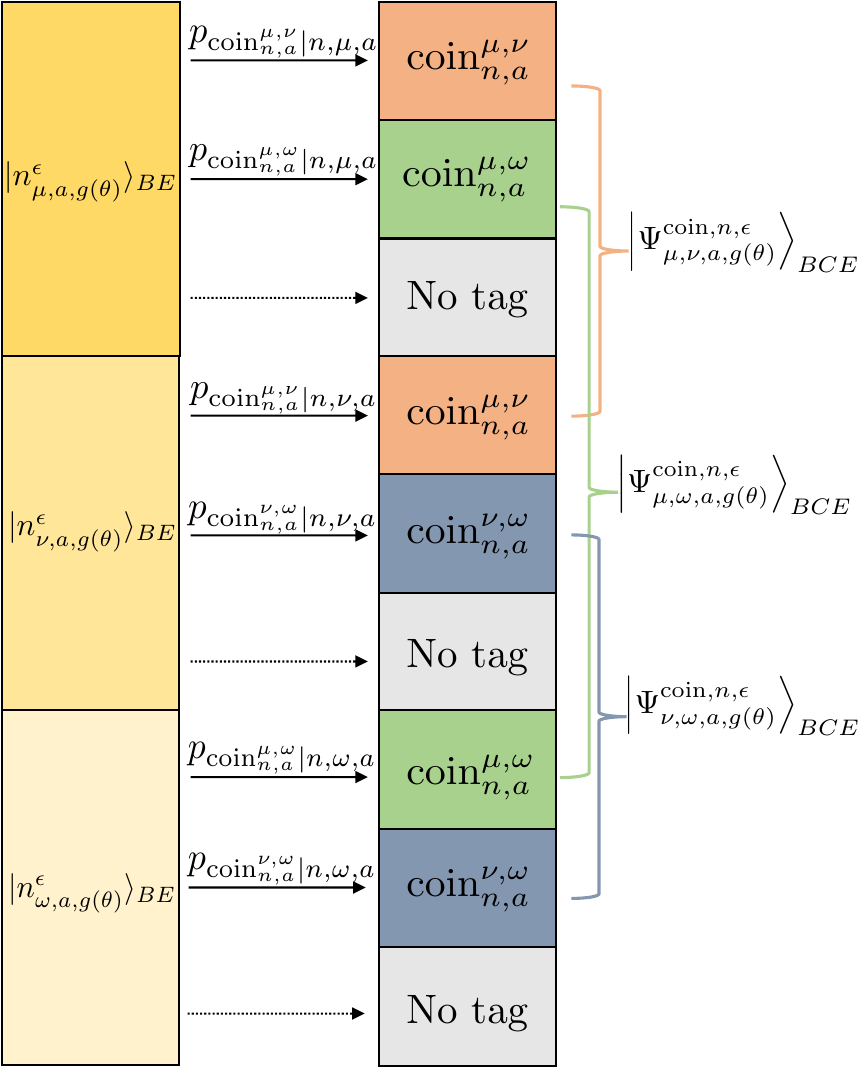}
\caption{\label{fig:tag}Graphical representation of the tagging process required to recreate the different symmetric quantum coins, as described in \cref{eq:a3}. To achieve this, the conditional tagging probabilities, $p_{\text{c}|n, \beta, a}$, must guarantee that the conditions expressed in \cref{eq:probsCoinRel} for the joint probabilities are satisfied. }
\end{figure}

We start by noticing that, from \cref{eq:states} we have that the actual protocol is equivalent to an alternative scenario in which, after selecting the settings $\beta$, $a$ and $\theta$ according to $g(\theta)$, Alice directly prepares the states $|n_{\beta,a, g(\theta)}^{\epsilon}\rangle_{BE}$ with probability $p^{\epsilon}_{n|\beta,a}$.
Moreover, we can further consider that she samples some of these emissions and assigns to them certain ``coin'' tags (in fact, Alice could assign more than one tag to each round). 
Precisely, let us denote by $p_{\text{c}|n, \beta, a}$ the conditional probability that Alice assigns the tag $\text{c}\in\set{\text{coin}^{\mu,\nu}_{n,a},\text{coin}^{\mu,\omega}_{n,a},\text{coin}^{\nu,\omega}_{n,a}}$ given that she emitted the state $|n_{\beta,a, g(\theta)}^{\epsilon}\rangle_{BE}$. Also let $p_{\text{c}, n, \beta, a}=p_{\text{c}|n, \beta, a}p_{n| \beta, a}^{\epsilon}p_{\beta}p_{a}$ denote the corresponding joint probability. Now, by selecting the tagging probabilities $p_{c|n,\beta,a}$ such that the constraints 
\begin{equation}\label{eq:probsCoinRel}
\begin{split}
    p_{\text{coin}^{\beta,\gamma}_{n,a}, n, \beta, a} &= p_{\text{coin}^{\beta,\gamma}_{n,a}, n, \gamma, a},\\
    p_{\text{coin}^{\beta,\gamma}_{n,a}, n, \zeta, a} &= 0,\qquad \zeta\notin\set{\beta,\gamma},
\end{split}
\end{equation}
are satisfied for the three possible coins $\set{\text{coin}^{\mu,\nu}_{n,a},\text{coin}^{\mu,\omega}_{n,a},\text{coin}^{\nu,\omega}_{n,a}}$, one can always define, for each coin, a second modified scenario in which, in these coin rounds, Alice directly prepares the states
\begin{equation}
\label{eq:a3}
\begin{aligned}
\ket{\Psi^{{\rm coin}, n, \epsilon}_{\beta,\gamma, a, g(\theta)}}_{BCE} &= \frac{1}{\sqrt{2}} \Big[\ket{0}_C\ket*{n_{\beta,a, g(\theta)}^{\epsilon}}_{BE} \\
&+ e^{i\phi_{c}}\ket{1}_C\ket*{n_{\gamma,a, g(\theta)}^{\epsilon}}_{BE}\Big],
\end{aligned}
\end{equation}
with certain probability $p_{\text{c}^{\beta,\gamma}_{n,a}}\leq 2\min[p_{n| \beta, a}^{\epsilon}p_{\beta}p_{a},p_{n| \gamma, a}^{\epsilon}p_{\gamma}p_{a}]$, where in \cref{eq:a3} we have included an arbitrary phase $\phi_{c}$. This tagging process is illustrated in \cref{fig:tag}.

Note that if Alice measures the coin system $C$ in \cref{eq:a3} in the computational basis before she sends system $BE$ to the channel, the previous alternative scenario is immediately recovered. 
Importantly, as we show below, for the coin rounds one can establish a relationship between the detection probabilities of any different intensities, say $Y^{n}_{\gamma,a}$ and $Y^{n}_{\beta,a}$. What is more, since the detection statistics of the transmitted states $\ket*{n_{\beta,a, g(\theta)}^{\epsilon}}_{BE}$ must be identical regardless of whether or not they have been sent in ``coin'' rounds, such statistical relation can be extended as well to all the transmitted rounds. Note that in the finite-key regime statistical fluctuations must be accounted for, and so defining the optimal tagging process can be a little bit more involved. 

Now, for the coin rounds it can be shown that \cite{lo_preskill}
\begin{equation}
\label{A3}
    1-2\Pr[X_C=-1]\leq \sqrt{Y^{n}_{\beta,a}Y^{n}_{\gamma,a}}+\sqrt{(1-Y^{n}_{\beta,a})(1-Y^{n}_{\gamma,a})},
\end{equation}
where $X_C$ represents the $X$ Pauli operator acting on the coin system $C$. Importantly, from \cref{eq:a3} we have that the LHS of the previous inequality can be computed as
\begin{equation}
\label{A4}
\begin{split}
    1-2\Pr[X_C=-1]&=\Re{e^{i\phi_c}\braket*{n_{\beta,a, g(\theta)}^{\epsilon}}{n_{\gamma,a, g(\theta)}^{\epsilon}}_{BE}}\\
    &=\big\vert\braket*{n_{\beta,a, g(\theta)}^{\epsilon}}{n_{\gamma,a, g(\theta)}^{\epsilon}}_{BE}\big\vert,
\end{split}
\end{equation}
where the last equality always holds for some phase $\phi_c$. Finally, by combining the previous two expressions, $i.e.$, \cref{A3,A4} one obtains
\begin{equation}
\label{eq:trace}
\begin{aligned}
&\big\vert\braket*{n_{\beta,a, g(\theta)}^{\epsilon}}{n_{\gamma,a, g(\theta)}^{\epsilon}}_{BE}\big\vert  \\
&\leq \sqrt{Y^{n}_{\beta,a}Y^{n}_{\gamma,a}}+\sqrt{(1-Y^{n}_{\beta,a})(1-Y^{n}_{\gamma,a})},
\end{aligned}
\end{equation}
from which~\cref{cs_text} follows directly \cite{pereira_2020}.

\subsection{Phase-error rate estimation}\label{app:coin}

Similar arguments to those used in the previous subsection can be applied as well to estimate an upper bound on the single photon phase-error rate $e_{\mu}^{{\rm ph}, 1}$. In particular, now our starting point is~\cref{eq:bas_dep_states}, which describes, for $\alpha\in\set{Z,X}$, the entangled states generated by Alice in a virtual protocol. The goal is to estimate the phase-error rate of the state $\ket*{\Psi_{\mu,Z, g(\theta)}^{n=1,\epsilon}}_{ABE}$ from the bit-error rate observed in those rounds in which Alice emits the state $\ket*{\Psi_{\mu,X, g(\theta)}^{n=1,\epsilon}}_{ABE}$ . 

To use the quantum-coin argument~\cite{lo_preskill} we consider again an additional equivalent coin protocol where some of the rounds of the virtual protocol in which Alice prepares $\ket*{\Psi_{\mu,Z, g(\theta)}^{n=1,\epsilon}}_{ABE}$ or $\ket*{\Psi_{\mu,X, g(\theta)}^{n=1,\epsilon}}_{ABE}$ are substituted by the preparation of the state
\begin{equation}
\label{eq:ideal_coin}
\begin{aligned}
\ket{\Phi^{{\rm coin},n=1, \epsilon}_{\mu, g(\theta)}}_{ABCE}=&\frac{1}{\sqrt{2}}\Big[\ket{0}_C\ket*{\Psi_{\mu,Z, g(\theta)}^{n=1,\epsilon}}_{ABE}\\
&+\ket{1}_C\ket*{\Psi_{\mu,X, g(\theta)}^{n=1,\epsilon}}_{ABE}\Big].
\end{aligned}
\end{equation}
Note that when $\epsilon\approx 0$, $\phi_{Z}\approx\phi_{X}$ (see \cref{eq:bas_dep_states}), and the source is close to a perfect phase-randomized source, then  these states satisfy $\abs*{\braket*{\Psi_{\mu,Z, g(\theta)}^{n=1,\epsilon}}{\Psi_{\mu,X, g(\theta)}^{n=1,\epsilon}}_{ABE}}^2\approx 1$ and so $\Pr[X_C=-1]\approx 0$. 

Let $p_{\Psi_{\alpha}^{n=1,\epsilon}}$ denote the probability of preparing the state $\ket*{\Psi_{\mu,\alpha, g(\theta)}^{n=1,\epsilon}}_{ABE}$ in the virtual protocol, which in general satisfies $p_{\Psi_{Z}^{n=1,\epsilon}}\neq p_{\Psi_{X}^{n=1,\epsilon}}$. Then, to make the equivalence between the virtual and the coin protocols precise, we consider again that, in every round of the virtual protocol, Alice assigns the tag ``coin'' to the emitted states $\ket*{\Psi_{\mu,\alpha, g(\theta)}^{n=1,\epsilon}}_{ABE}$ with certain probability $p_{\text{coin}|\Psi_{\alpha}^{n=1,\epsilon}}$. Importantly, this is done such that those rounds that are tagged as ``coin'' are indistinguishable from the emission of the symmetric quantum coin defined in Eq.~\eqref{eq:ideal_coin}. This can be guaranteed if the following condition is satisfied
\begin{equation}
    p_{\text{coin}|\Psi_{Z}^{n=1,\epsilon}} p_{\Psi_{Z}^{n=1,\epsilon}} = p_{\text{coin}|\Psi_{X}^{n=1,\epsilon}} p_{\Psi_{X}^{n=1,\epsilon}}.
\end{equation}
That is, in this fictitious scenario, Alice effectively prepares the coin state $\ket*{\Phi^{n=1, \epsilon}_{{\rm coin},\mu, g(\theta)}}_{ABCE}$ with certain probability $p_{\text{coin}}\leq 2\min[p_{\Psi_{Z}^{n=1,\epsilon}},p_{\Psi_{X}^{n=1,\epsilon}}]$. Note that the latter bound implies that $p_{\text{coin}|\Psi_{\alpha}^{n=1,\epsilon}}\leq \min[p_{\Psi_{Z}^{n=1,\epsilon}},p_{\Psi_{X}^{n=1,\epsilon}}]/p_{\Psi_{\alpha}^{n=1,\epsilon}}$.

Once again, since we are considering here the asymptotic-key regime, the phase-error rate of the sifted key is essentially identical to the phase-error rate estimated from the coin rounds, and so we can simply set the probabilities $p_{\text{coin}|\Psi^{n=1,\epsilon}_{\alpha}}$ to a very small value (as long as the previous relation is satisfied).

Finally, as shown in \cite{GLLP}, one can use \cref{qwe} to bound the deviation between the phase-error rate in the $Z$ basis and the $X$-basis bit-error rate within the detected coin rounds. This bound requires to know the quantum-coin imbalance parameter $\Delta^{n}_{\beta}$ ---or, equivalently, the probability $\Pr[X_C=-1]$--- but now within the detected rounds. For this, we consider a very pessimistic worst-case scenario in which all of the events that satisfy $X_C=-1$ are detected, leading to \cref{eq:basis_dependence}. The drawback of this approach is that it results in a poor performance of the protocol at long distances (when $Y^{n,\rm{L}}_{\beta, \rm{coin}}\sim 1-\sqrt{F_{\beta}^{n,\epsilon,{\rm L}}}$) unless the bound on the fidelity $F_{\beta}^{n=1,\epsilon}\equiv\abs*{\braket*{\Psi_{\beta,Z, g(\theta)}^{n=1,\epsilon}}{\Psi_{\beta,X, g(\theta)}^{n=1,\epsilon}}_{ABE}}^2$ is very close to one.

\section{Results on matrix perturbation theory}\label{app:use}
In this Appendix, we present three results from the field of matrix perturbation theory that we use in the main text (see $e.g.$ \cref{eq:bound_main} and \cref{eq:fide_bound_2}). For more details, the reader is refereed to \cite{Nahar, phase1}.

{\bf Result 1 :} Let $\mathcal{H}$ be a Hilbert space. Given $\rho,\sigma \in$ $\mathrm{D}(\mathcal{H})$, where $ \mathrm{D}(\mathcal{H})$ represents the density operators acting on $\mathcal{H}$, such that $\|\sigma-\rho\|_1 \leq 2 \kappa$, then
\begin{equation}
\label{eq:eigen_bounds}
\left|\lambda_n(\rho)-\lambda_n(\sigma)\right| \leq \kappa,
\end{equation}
where $\|\cdot\|_1$ represents the trace norm and $\lambda_n(\rho)$ is the $n$-th highest eigenvalue of $\rho$. 

{\bf Result 2:} Let $\rho$, $\sigma$ be two density operators satisfying $\|\sigma-\rho\|_1 \leq 2 \kappa$. Let $\delta_n=\min \left\{\lambda_n(\sigma)-\lambda_{n-1}(\sigma), \lambda_{n+1}(\sigma)-\lambda_n(\sigma)\right\}-\kappa$, where $\lambda_n(\sigma)$ is the $n$-th highest eigenvalue of $\sigma$, then
\begin{equation}
\label{eq:D-K}
F\big(\ket{n(\rho)},\ket{n(\sigma)}\big)\geq 1- \frac{\kappa^{2}}{\delta_{n}^{2}}:=1-\gamma_{n},
\end{equation}
with $\ket{n(\rho)}$ and $\ket{n(\sigma)}$ being the $n$-th eigenvectors of $\rho$ and $\sigma$, respectively. This result is a variation of the Davis-Kahan (D-K) theorem \cite{DK}.

In addition, we have that the fidelity between two density matrices can be lower bounded by the fidelity between any of their purifications. In particular, if we consider the states given by \cref{eq:ideal,eq:states}, it is straightforward to show that $F(\rho^{\beta,a}_{g(\theta),BE},\rho^{\epsilon,\beta,a}_{g(\theta),BE})\geq 1-\epsilon$. Having established this bound, we can use the Fuchs-van de Graaf inequality, which is defined for two arbitrary density matrices $\rho$ and $\sigma$ as \cite{nielsen}
\begin{equation}
1-\sqrt{F(\rho, \sigma)} \leq \frac{1}{2}\|\rho-\sigma\|_{1} \leq \sqrt{1-F(\rho, \sigma)},
\end{equation}
to find a bound on the desired trace norm. This way, we have that
\begin{equation}\label{eq:epsilon}
\begin{aligned}
&\norm{\rho^{\beta,a}_{g(\theta),BE}-\rho^{\epsilon,\beta,a}_{g(\theta),BE}}_1 \\
&\leq 2\sqrt{1-F(\rho^{\beta,a}_{g(\theta),BE},\rho^{\epsilon,\beta,a}_{g(\theta),BE})}\leq2\sqrt{\epsilon}=2\kappa,
\end{aligned}
\end{equation}
which implies that $\kappa=\sqrt{\epsilon}$.

{\bf Result 3:} Let $\rho$ be a density matrix, and let $\rho^{\prime}=\Pi \rho \Pi/\operatorname{Tr}[\Pi \rho \Pi]$, where $\Pi$ denotes a projector onto a subspace of dimension $M+1$. Then,
\begin{equation}
\begin{aligned}
F\left(\rho, \rho^{\prime}\right)=&\operatorname{Tr}\left[\sqrt{\sqrt{\rho} \rho^{\prime} \sqrt{\rho}}\right]^2\\
&=\frac{\operatorname{Tr}[\sqrt{\sqrt{\rho} \Pi \rho \Pi \sqrt{\rho}}]^2}{\operatorname{Tr}[\Pi \rho \Pi]}=\operatorname{Tr}[\Pi \rho \Pi]= \sum_{n=0}^{M}  \lambda_n
\end{aligned}
\end{equation}
where $\lambda_n$ are the eigenvalues of $\Pi \rho \Pi$.

\section{Linear Cauchy-Schwarz constraint for parameter estimation}\label{app:linear_CS}

To enable the use of linear programming for the decoy-state parameter estimation procedure, a linear version of the CS constraints given by Eq.~(\ref{cs_text}) is needed. As shown in \cite{Zapatero_2021}, such linearized versions for yield estimation (and similarly for the bit-error rate estimation) can be expressed as
\begin{widetext}
\begin{equation}
\label{linearyield}
\begin{aligned}
&G_{-}\left(\tilde{Y}^{n}_{\zeta, a}, \vert\braket*{n_{\zeta,a,g(\theta)}^{\epsilon}}{n_{\gamma,a,g(\theta)}^{\epsilon}}_{BE}\vert^2\right)+G_{-}^{\prime}\left(\tilde{Y}^{n}_{\zeta, a}, \vert\braket*{n_{\zeta,a,g(\theta)}^{\epsilon}}{n_{\gamma,a,g(\theta)}^{\epsilon}}_{BE}\vert^2\right)\left(Y^{n}_{\zeta, a}-\tilde{Y}^{n}_{\zeta, a}\right) \leq Y^{n}_{\gamma, a}  \\
&\leq G_{+}\left(\tilde{Y}^{n}_{\zeta, a}, \vert\braket*{n_{\zeta,a,g(\theta)}^{\epsilon}}{n_{\gamma,a,g(\theta)}^{\epsilon}}_{BE}\vert^2\right)+G_{+}^{\prime}\left(\tilde{Y}^{n}_{\zeta, a}, \vert\braket*{n_{\zeta,a,g(\theta)}^{\epsilon}}{n_{\gamma,a,g(\theta)}^{\epsilon}}_{BE}\vert^2\right)\left(Y^{n}_{\zeta, a}-\tilde{Y}^{n}_{\zeta, a}\right),
\end{aligned}
\end{equation} 
\end{widetext}
where  $\tilde{Y}^{n}_{\beta, a}$ represents the reference parameter of the linear approximation. Finally, the functions $G_{\pm}^{\prime}$ are defined as

\begin{equation}\label{brussels}
\begin{aligned}
&G_{-}^{\prime}(y, z)=\begin{cases}g_{-}^{\prime}(y, z) & \text { if } y>1-z \\
0 & \text { otherwise }\end{cases},
\\
&G_{+}^{\prime}(y, z)= \begin{cases}g_{+}^{\prime}(y, z) & \text { if } y<z \\
0 & \text { otherwise }\end{cases},
\end{aligned}
\end{equation}
with $g_{\pm}^{\prime}(y, z)=-1+2 z \pm(1-2 y) \sqrt{z(1-z) / y(1-y)}$ being the derivative of the function $g_{\pm}(y,z)$ given by \cref{eq:smallg} with respect to $y$.

Importantly, to make the linear constraints as tight as possible, the reference parameters $\tilde{Y}^{n}_{\beta, a}$ can be optimized numerically, as any reference value offers a valid bound. Note that prior studies that employ these constraints \cite{ Zapatero_2021,sixto2022}, did not perform such optimization. Instead, they set the reference yield (reference error) based on the expected behavior of the channel. To fine-tune this technique, we employ the optimal values of $Y^{n}_{\beta, a}$ outputted by the linear programs at a given distance point as the reference values for the next distance point. For the initial point, that is at zero distance, we establish a close-to-optimal reference value by executing the linear program multiple times in a Monte Carlo fashion with different reference parameters and selecting the highest output. 

As a side remark, we note that if needed, one could always add more constraints to the linear programs by considering multiple reference values for each quantity to be bounded.

\section{Fidelity bound using semidefinite programming.}\label{sec:inner_SDP}

Using the CS type constraints requires to calculate a bound on the inner product between two states with different intensity settings, namely  $\vert\braket*{n_{\gamma,a,g(\theta)}^{\epsilon}}{n_{\zeta,a,g(\theta)}^{\epsilon}}_{BE}\vert$.

As shown in \cref{eq:D-K}, if one considers the fidelity between an ideal state (with no side channel) and an imperfect state (with side channel) with the same intensity setting, we have that
\begin{equation}
\label{eq:bound}
    \vert\braket*{n_{\beta,a,g(\theta)}^{\epsilon}}{n_{\beta,a,g(\theta)}}_{BE}\vert^2 \geq 1- \frac{\kappa^{2}}{\delta_{n,\beta}^{2}}:= 1-\gamma_{n, \beta}.
\end{equation}
Importantly, in \cite{Guille_framework} it is shown that \cref{eq:bound} implies that one can assume the states $\ket*{n_{\beta,a,g(\theta)}^{\epsilon}}_{BE}$ to have the form
\begin{equation}
\label{eq:descom}
\begin{aligned}
    &\ket*{n_{\beta,a,g(\theta)}^{\epsilon}}_{BE}  \\
    &=\sqrt{1-\gamma_{n, \beta}} \ket*{n_{\beta,a,g(\theta)}}_{BE} + \sqrt{\gamma_{n, \beta}} \ket*{n_{\beta,a,g(\theta)}^{\perp}}_{BE},
\end{aligned}
\end{equation}
where $\ket*{n_{\beta, a,g(\theta)}^{\perp}}_{BE}$ is some state orthogonal to $\ket*{n_{\beta, a,g(\theta)}}_{BE}$. 

To obtain a lower bound on $\vert\braket*{n_{\gamma,a,g(\theta)}^{\epsilon}}{n_{\zeta,a,g(\theta)}^{\epsilon}}_{BE}\vert$, we employ semidefinite programming (SDP). For this, from \cref{eq:descom} we have that
\begin{equation}
\begin{aligned}
\label{eq:Re_bound}
&\vert\braket*{n_{\gamma,a,g(\theta)}^{\epsilon}}{n_{\zeta,a,g(\theta)}^{\epsilon}}_{BE}\vert \geq \Re \braket*{n_{\gamma,a,g(\theta)}^{\epsilon}}{n_{\zeta,a,g(\theta)}^{\epsilon}}_{BE} \\
&= \sqrt{1-\gamma_{n, \gamma}}\sqrt{1-\gamma_{n, \zeta}} \Re \braket*{n_{\gamma,a,g(\theta)}}{n_{\zeta,a,g(\theta)}}_{BE} \\
&+\sqrt{1-\gamma_{n, \gamma}} \sqrt{\gamma_{n, \zeta}} \Re \braket*{n_{\gamma,a,g(\theta)}}{n_{\zeta,a,g(\theta)}^{\perp}}_{BE} \\
&+\sqrt{1-\gamma_{n, \zeta}} \sqrt{\gamma_{n, \gamma}}\Re \braket*{n_{\gamma,a,g(\theta)}^{\perp}}{n_{\zeta,a,g(\theta)}}_{BE}\\
&+ \sqrt{\gamma_{n, \gamma}}\sqrt{\gamma_{n, \zeta}} \Re \braket*{n_{\gamma,a,g(\theta)}^{\perp}}{n_{\zeta,a,g(\theta)}^{\perp}}_{BE}=: F.
\end{aligned}
\end{equation}

Next, we define $G$ as the Gram matrix of the vector set $\{\ket*{n_{\zeta,a,g(\theta)}}_{BE}, \ket*{n_{\gamma,a,g(\theta)}}_{BE}, \ket*{n_{\zeta,a,g(\theta)}^\perp}_{BE}, \ket*{n_{\gamma,a,g(\theta)}^\perp}_{BE} \}$. That is, $G[2,1] = \braket*{n_{\gamma,a,g(\theta)}}{n_{\zeta,a,g(\theta)}}_{BE}$ refers to the inner product between the second and first vectors of the list and so on. This way, we can rewrite \cref{eq:Re_bound} as
\begin{equation}
\begin{aligned}
\Re \braket*{n_{\gamma,a,g(\theta)}^{\epsilon}}{n_{\zeta,a,g(\theta)}^{\epsilon}}_{BE} &\geq\sqrt{1-\gamma_{n, \gamma}}\sqrt{1-\gamma_{n, \zeta}} \Re G[2,1] \\
&+\sqrt{1-\gamma_{n, \gamma}} \sqrt{\gamma_{n, \zeta}} \Re G[2,3]\\
&+\sqrt{1-\gamma_{n, \zeta}} \sqrt{\gamma_{n, \gamma}} \Re G[4,1] \\
&+\sqrt{\gamma_{n, \gamma}}\sqrt{\gamma_{n, \zeta}} \Re G[4,3]=:F,
\end{aligned}
\end{equation}
where $\Re G[i,j] = (G[i,j]+G[j,i])/2$, since $G$ is Hermitian. Also, we can express all the information we have as constraints on elements of $G$, $i.e.$,
\begin{equation}
    \begin{aligned}
    &G[j,j] = 1, \quad \forall j \in \{1,2,3,4\}, \\
    &G[j,j+2] = 0, \quad \forall j \in \{1,2\}, \\
    &G[2,1] = \braket{n_{\gamma,a,g(\theta)}}{n_{\zeta,a,g(\theta)}}_{BE},
    \end{aligned}
\end{equation}
where $\braket{n_{\gamma,a,g(\theta)}}{n_{\zeta,a,g(\theta)}}_{BE}$ is known. Putting all together, we find that a lower bound on $\Re \braket*{n_{\gamma,a, g(\theta)}^{\epsilon}}{n_{\zeta, a,g(\theta)}^{\epsilon}}_{BE}$ can be obtanied by solving the dual of the following SDP problem
\begin{equation}
\label{eq:opt_prob}
\begin{aligned}
\min_G \textrm{  }&F \\
\textrm{s.t.  }& G[j,j] = 1, \quad \forall j \in \{1,2,3,4\}, \\
&G[j,j+2] = 0, \quad \forall j \in \{1,2\}, \\
&G[2,1] = \braket{n_{\gamma,a,g(\theta)}}{n_{\zeta,a,g(\theta)}}_{BE},  \\
&G \geq 0. 
\end{aligned}
\end{equation}

This is because any solution to the dual problem is guaranteed to be lower or equal than the optimal solution of the primal problem, which is itself guaranteed to be lower or equal than the actual value of $\Re \braket*{n_{\gamma,a,g(\theta)}^{\epsilon}}{n_{\zeta,a,g(\theta)}^{\epsilon}}_{BE}$. Importantly, as shown in \cref{eq:Re_bound}, this provides a valid bound for $\vert\braket*{n_{\gamma,a,g(\theta)}^{\epsilon}}{n_{\zeta,a,g(\theta)}^{\epsilon}}_{BE}\vert.$

\section{Channel model}\label{app:channel_model}

\begin{figure}[htpb]
\centering\includegraphics [width= 8.6cm, height=3.4cm] {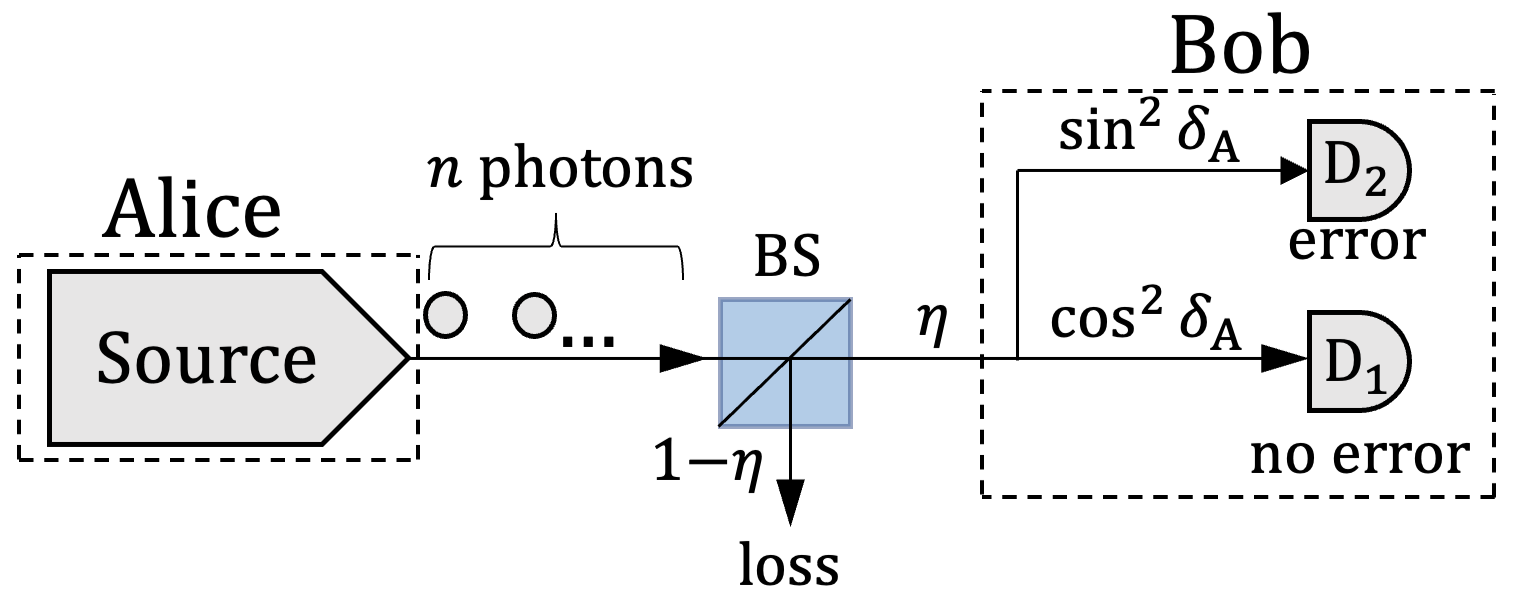}
\caption{\label{fig:channel}Pictorial representation of the channel model employed in the simulations presented in the main text. It is based on that presented in \cite{mali}, where  $\eta$ is the overall system efficiency, encompassing both channel loss and the finite detection efficiency of Bob's detectors, while $\delta_{A}$ signifies the polarization misalignment introduced by the channel. BS denotes a beamsplitter. On his side, Bob employs an active BB84 receiver equipped with two detectors, labeled by $D_1$ and $D_2$. If both detectors click, the outcome is randomly assigned to one of them. }
\end{figure}

In this Appendix we provide the experimental inputs of the linear programs presented in the main text, namely, the parameters $Q_{\beta,  a}$, and $E_{\beta,  a}Q_{\beta,  a}$ according to a standard channel model illustrated in \cref{fig:channel}. For this purpose, let $\eta_{\mathrm{det}}=0.65$ represent the common detection efficiency of Bob’s detectors and let $p_{\mathrm{d}}=7.2\times 10^{-8}$ indicate the dark count probability of each of Bob’s detectors --- both values reported in \cite{data} --- and let $\eta_{\mathrm{ch}} = 10^{-\alpha_{\text{dB}} L / 10}$ denote the transmittance of the quantum channel, where $\alpha_{\text{dB}}=0.2$ (dB/km) is its attenuation coefficient and $L$ (km) is the distance. Finally $\delta_{\mathrm{A}}=0.08$ signifies the misalignment occurring in the channel. 

Calculations of the parameters $Q_{\beta,  a}$ and $E_{\beta, a}$ \cite{mali} result in,
\begin{equation}
Q_{\beta,  a}=1-\left(1-p_{\mathrm{d}}\right)^{2} e^{-\eta \beta}, 
\end{equation}
and
\begin{equation}
\begin{aligned}
E_{\beta, a}Q_{\beta,  a}=&\frac{p_{\mathrm{d}}^{2}}{2}+p_{\mathrm{d}}\left(1-p_{\mathrm{d}}\right) \left(1+h_{\eta, \beta, \delta_{\mathrm{A}}}\right)\\
&+\left(1-p_{\mathrm{d}}\right)^{2} \left(\frac{1}{2}+h_{\eta, \beta, \delta_{\mathrm{A}}}-\frac{1}{2} e^{-\eta \beta}\right),
\end{aligned}
\end{equation}
where $\eta=\eta_{\mathrm{det}}\eta_{\mathrm{ch}}$ stands for the overall system attenuation, and 
\begin{equation}
h_{\eta, \beta, \delta_{\mathrm{A}}}=\frac{1}{2}\left(e^{-\eta \beta \cos ^{2} \delta_{\mathrm{A}}}-e^{-\eta \beta \sin ^{2} \delta_{\mathrm{A}}}\right).
\end{equation}  

\bibliography{refs}

\end{document}